\documentclass[aps,twocolumn,superscriptaddress,showpacs]{revtex4-1}

\usepackage{amsmath}
\usepackage{amsfonts}
\usepackage{amssymb}
\usepackage{bm}
\usepackage{graphicx}

\usepackage{color}

\begin{document}

\title{Properties of Mesons in a Strong Magnetic Field}

\author{Rui Zhang }
\affiliation{Department of Physics and State Key Laboratory of Nuclear Physics and Technology, Peking University, Beijing 100871, China}

\author{Wei-jie Fu }
\affiliation{Institut f\"{u}r Theoretische Physik, Universit\"{a}t Heidelberg, Philosophenweg 16, 69120 Heidelberg, Germany}

\author{Yu-xin Liu }
\email[Corresponding author: ]{yxliu@pku.edu.cn}
\affiliation{Department of Physics and State Key Laboratory of Nuclear Physics and Technology, Peking University, Beijing 100871, China}
\affiliation{Collaborative Innovation Center of Quantum Matter, Beijing 100871, China}
\affiliation{Center for High Energy Physics, Peking University, Beijing 100871, China}

\date{\today}

\begin{abstract}
By extending the $\Phi$--derivable approach in Nambu--Jona-Lasinio model to finite magnetic field we calculate the properties of pion, $\sigma$ and $\rho$ mesons in a magnetic field at finite temperature in not only the quark--antiquark bound state scheme but also the pion--pion scattering resonant state scenario.
Our calculation results manifest that the masses of $\pi^{0}$ and $\sigma$ meson can be nearly degenerate at the pseudo-critical temperature which increases with increasing the magnetic field strength, and the $\pi^{\pm}$ mass ascends suddenly at almost the same critical temperature.
While the $\rho$ mesons' masses decrease with the temperature but increase with the magnetic field strength.
We also check the Gell-Mann--Oakes--Renner relation and find that the relation can be  violated  obviously with increasing the temperature,
and the effect of the magnetic field becomes pronounced around the critical temperature.
With different criteria, we analyze the effect of the magnetic field on the chiral phase transition
and find that the pseudo-critical temperature of the chiral phase cross, $T_{c}^{\chi}$, is always enhanced by the magnetic field.
Moreover our calculations indicate that the $\rho$ mesons will get melted as the chiral symmetry has not yet been restored,
but the $\sigma$ meson does not disassociate even at very high temperature.
Particularly, it is the first to show that there does not exist vector meson condensate in the QCD vacuum in the pion--pion scattering scheme.
\end{abstract}

\pacs{12.38.Mh, 25.75.Nq,11.10.Wx, 12.39.Ki}

\maketitle



\section{Introduction}

The properties of strong interaction matter (QCD matter) have attracted great
attentions in the past years, and plenty of theoretical and
experimental results are obtained (see, for example, Refs.~\cite{Gyulassy:2005NPA,sQGP-OriginalExp,Wambach:2009Review,Gupta:2011SCI,Fukushima:2012JPG,Fukushima:2011RPP,Stephanov:2007POSL,Jiang:2013PRD,Xin:2014PRD,Qin:2011PRL,Liu:2011AIPCP,Fischer:201134,Pawlowski:2013PRD,Philipsen:2013,Bazavov:2012PRD,Borsanyi:2010JHEP,Liu:2011PRD,Gavai:2011PLB,DElia:20101PRD,Bali:20124,Philipsen:2014PRL,Blaizot:2007JPG,Kharzeev:2008NPA,Chernodub:2010PRD,Adler:2003PRL,Adamczyk:2014PRL,Adamczyk:2015PRL,Aamodt:2010PRL}).
The complicated phase structure of the matter provides rich information on the property of strong interaction at finite temperature and/or  density,
and may shed light on fundamental understanding for some basic
problems, {\it e.g.}, the origin of most mass of visible matter and the evolution of the
early universe matter. The chiral phase transition which is expected to occur in
ultrarelativistic heavy-ion collisions~\cite{Gyulassy:2005NPA,sQGP-OriginalExp,Blaizot:2007JPG} and/or in the interior of the
highly compact stars~\cite{Alford:2006NAT,Fu:2008PRL,Weber:2014PRC} is one of the
significant issues for the research of QCD matter.
An important method to extract the information of the chiral phase transition is analyzing the variation of hadrons' properties in the medium at finite temperature or/and density (chemical potential), even in finite magnetic field, compared with those at zero temperature, zero chemical potential and zero magnetic field strength~\cite{Chernodub:2010PRD,Quack:1995PLB,Cassing:1999Rev,Brambill:2011Rev,Oertel:2000NPA,Laermann:2001EPJC,Costa:2005PRD,Naruki:2006PRL,Hansen:2006PRD,Fu:2009PRD,Blaschke:2011PRD,Wang:2013DSE,Fayazbakhsh:2013PRD}.
Among hadrons, mesons are more important than baryons at present stage, because the former is more sensitive to the change of the surroundings and related to the chiral phase transition more directly~\cite{Wang:2013DSE,Wang:2012DSE}.
Since it has been known that, when the early universe experienced the cosmological electro-weak phase transition,
the strength of the magnetic field may reach up to $eB \approx 200m_{\pi}^{2}$~\cite{Vachaspati:1991nm}.
In heavy-ion collision experiments, magnetic field produced at the early stage of non-central collisions can be the order of
$eB \approx 0.1 m_{\pi}^{2}$ for SPS, $eB \approx m_{\pi}^{2}$ for RHIC and
$eB \approx 15 m_{\pi}^{2}$ for LHC~\cite{Skokov:2009qp}.
Even though it is weaker than that at the early stage of the universe,
the magnetic field produced in RHIC or/and LHC has been strong enough to influence the strong interaction matter significantly~\cite{Kharzeev:2008NPA}.
Effects of magnetic field on the masses of hadrons and weak decay constant of neutral pion have then been investigated~\cite{Andreichikov:2013Hamiltonian,Hidaka:2013lattice,Fayazbakhsh:2013PRD,Huang:2015PRD,Taya:2015PRD,Huang:2015Pre}.
Since they are significant to check the validity of the Vafa-Witten theorem for the QCD vacuum, the variation behaviors of $\rho$ meson masses with respect to magnetic field strength
have also been studied~\cite{Andreichikov:2013Hamiltonian,Hidaka:2013lattice,Taya:2015PRD,Huang:2015PRD,Huang:2015Pre}.
However, the temperature and magnetic field strength dependence of the $\rho$ meson mass and
the existence of $\rho$ meson condensate in very strong magnetic filed at high temperature are still under debate (see, {\it e.g.}, Refs.~\cite{Andreichikov:2013Hamiltonian,Hidaka:2013lattice,Taya:2015PRD} and Refs.~\cite{Huang:2015PRD,Huang:2015Pre}).
In this paper, we consider further the 2 flavor QCD matter and the properties of mesons, including not only the pseudoscalar neutral pion and scalar $\sigma$ meson but also the charged pion and $\rho$ mesons, and show the impossibility of the $\rho$ meson condensate by analyzing the variation behavior of the masses and the width of the mass pole.

In general point of view, researches on the magnetic field effect have been carried out much more widely and many progresses have been made, {\it e.g.}, there may exist a chiral magnetic effect,
which demonstrates that imbalanced chirality in a magnetic field can
induce a current along the magnetic field and results in a separation of electric charges~\cite{Kharzeev:2008NPA,Adamczyk:2015PRL,Fukushima:2008xe,Buividovich:2009,Fukushima:2010PRD,Fu:2011CTP,Kharzeev:2011PRL,Burnier:2011PRL,RHIC:2014PRL,RHIC:2015PRL}.
The researches have simultaneously arisen some open questions.
One of them is whether there exists a magnetic catalysis, which says that the quark condensate would be enhanced with increasing the magnetic field~\cite{Kawati:1983aq,Suganuma:1990nn,Miransky:2002rp}.
A direct consequence of the magnetic catalysis is that the critical temperature of the chiral phase transition increases monotonically with the increasing of the magnetic field~\cite{Miransky:2002rp,Smilga:1997PLB,DElia:20101PRD,Kabat:2002PRD,Chernodub:2011PRL}.
However latter lattice QCD calculations show an opposite behavior, called
inverse magnetic catalysis, which presents a decreasing or non-monotonous behavior of
the critical temperature with increasing magnetic field~\cite{Bali:20124,Ilgenfritz:2013PRD}.
Lots of works have then been accomplished in order to determine whether the magnetic catalysis or the inverse
catalysis is correct (see, for instance, Refs.~\cite{Bali:20124,Ayala:2012PRD,Fukushima:2013PRL,Kojo:2013PLB,Bruckmann:2013JHEP,Chao:2013PRD,Fraga:2014PLB,Andersen:2014JHEP,Ruggieri:2014PLB,Fischer:2014DSE,Ferreira:2014PRD,Huang:2014PRD,Zong:2015PRD,Braun:2014,Pawlowski:2015}),
but it is still a puzzle (for a review, see, Ref.~\cite{Andersen:2014}).
Since mesons carry lots of information on the dynamical chiral symmetry breaking (DCSB) and restoration (in particular, pions),
and the proposed enhancement of the meson or quark--antiquark pair condensate induced by the strong magnetic field~\cite{Kabat:2002PRD,Ayala:2012PRD,Chernodub:2011PRL} may be a signature of the magnetic catalysis,
it is then expected that studies of meson properties in a strong magnetic field in this paper
would shed light on this open question.

If we consider the charged mesons ($\pi^\pm$ and $\rho^\pm$) as point
particles in an external magnetic field $B$, which is along the $z$
direction, the energy level of the particle with mass $m$ and spin
$s$ can be expressed as~\cite{Chernodub:2010PRD}:
\begin{equation}
\varepsilon^2_{n,s_z}(p_z)=p^2_z+(2n-2s_z+1)eB+m^2,
\end{equation}
where $n$ denotes the order number of the Landau level. As mentioned in
Ref.~\cite{Chernodub:2010PRD}, for pion $s_{z}^{}=0$, and for $\rho$ meson
$s_{z}^{}=1$, then the ground state mass of the charged pion and $\rho$
meson are given as:
\begin{equation}
m^2_{\pi^\pm}(B)=m^2_{\pi^\pm}+eB,\label{mpi}
\end{equation}
\begin{equation}
m^2_{\rho^\pm}(B)=m^2_{\rho^\pm}-eB,
\end{equation}
where $m^2_{\pi^\pm}$ and $m^2_{\rho^\pm}$ are the zero-field vacuum
masses of the $\pi^\pm$ and $\rho^\pm$. However, mesons are not
point particles. We can not ignore the contribution of the internal
quark structure of the particle to its mass, so in this paper we
will calculate the mesons' masses from the internal quark--antiquark
contribution and at the same time make correction on the point particle approximation.

It has been known that the Nambu--Jona-Lasinio (NJL) model is a QCD-inspired model~\cite{Klevansky:1992Review,Hatsuda:1994Review,Bijnens:1996Review,Buballa:2003Review},
which demonstrates the effects of chiral symmetry and its breaking well and, in turn, can describe
the meson properties at finite temperature successfully.
Meanwhile, the $\Phi$--derivable approximation~\cite{Lee:1960PR,Luttinger:1960PR,Baym:1961PR,Blaizot:2004NPA} has been known as a non-perturbative approach to the quantum field theory~\cite{Blaizot:2011PLB}, at least a two-particle-irreducible effective action formalism~\cite{CJT:1974PRD}. In this paper, we take then the NJL model with an extension of the $\Phi$-derivable scheme to finite magnetic field to calculate the meson properties in the conventional view that the mesons are quark and antiquark bound states. However, the $\sigma$-meson may not be a simple quark--antiquark bound state (see, {\it e.g.}, Refs.~\cite{Chen:2007PRD,Fischer:2012PLB}) but a resonant state of pion--pion scattering, and so does the $\rho$ meson (for reviews, see {\it e.g.},  Refs.~\cite{Ananthanarayan:2001PR,Bugg:2004PR,Nakamura:2010Review,Oller:2012PRD}).
We will then also study the mesons' properties by analyzing the $\pi$--$\pi$ scattering lengths and the resonant states.

This paper is organized as follows. In Sec. 2, we briefly reiterate the scheme
of describing the mesons in view of their internal quark--antiquark structure in the 2 flavor NJL model with the $\Phi$-derivable scheme at finite temperature but zero magnetic field.
In Sec. 3, we extend the formulation to the case of finite magnetic field.
In Sec. 4, numerical results and discussions of the dependence of meson properties
on the magnetic field are presented.
In Sec. 5, we re-calculate the masses and the widths of the $\sigma$ and $\rho$ mesons
in view of the $\pi$--$\pi$ scattering resonant states to reanalyze the effect of the magnetic field in an alternative scheme.
In Sec. 6, we will give a summary and some remarks.

\section{Meson properties in the NJL model without magnetic field}

We begin with the NJL Lagrangian:
\begin{eqnarray}
\mathcal{L} & = & \overline{\psi}(i\partial \!\!\!/ - m_{0}^{}) \psi + g_{s}^{} [(\overline{\psi}\psi)^{2} + (\overline{\psi} i \gamma_{5}^{} \overrightarrow{\tau}\psi)^{2} ] \notag \\
&& - g_{v}^{} (\overline{\psi} \gamma^{\mu} \overrightarrow{\tau} \psi )^{2} ,
\end{eqnarray}
where $\psi$ and $\overline{\psi}$ denote a quark and an antiquark field with $N_{f}^{}$ flavors and $N_{c}$ colors, $m_{0}^{}$ is the bare quark mass, $\tau^{i}(i=1,2,3)$ are the Pauli matrices in flavor space. The effective four fermion interaction constant for scalar and pseudoscalar channels is $g_{s}^{}$, and that for the vector channel is $g_{v}^{}$. In this paper we always treat $N_{f}^{}$ and $N_{c}$ as constants, $N_{f}^{}=2$ and $N_{c}=3$.

To make use of the $\Phi$-derivable theory in practical calculation, we follow exactly the scheme
described in Refs.~\cite{Blaizot:2004NPA,Fu:2013PRD}. We skip then the complicated derivations
and only list some important results as the follows.
At first, the constituent quark mass can be derived from the gap equation
\begin{equation}
M=m_{0}^{} + \Sigma \, ,  \label{gap-Eq}
\end{equation}
where $\Sigma$ is the quark self-energy function. In lowest order approximation, it reads
\begin{equation}
\Sigma = 2 {g_{s}^{}} \int\frac{d^4q}{(2\pi)^4}\mathrm{tr}(iS(q)) \, , \label{sigma}
\end{equation}
with full quark propagator $S(q)=1/(q \!\!\!/ - M)$.
The trace notation, tr, acts in the Dirac, flavor and color spaces.
The quark condensate $\langle\overline{q}q\rangle$ is defined as
\begin{equation}    \label{Condensate-field}
\langle \bar{q} q \rangle = \phi=-\int\frac{d^4q}{(2\pi)^4}\mathrm{tr}(iS(q)).
\end{equation}
To show the flavor dependence of the condensate, the trace in Eq.~(\ref{Condensate-field}) does not include that in flavor space usually. Comparing with Eqs.~(\ref{sigma}) and (\ref{gap-Eq}) we can get
\begin{equation}
\phi=-\frac{M - m_{0}}{4{g_{s}^{}} } \, . \label{phi}
\end{equation}

In view that mesons are bound states of a quark and an antiquark,
the meson propagators can be represented in terms of its ``polarization function" $\Pi_{\alpha}(p)$ as
\begin{eqnarray}
&&D_\sigma(p)=\frac{2g_s}{1-2g_s\Pi_\sigma(p)} \, , \\
&&D_\pi(p)=\frac{2g_s}{1-2g_s\Pi_\pi(p)}\, , \\
&&D_\rho(p)=\frac{2g_v}{1-2g_v\Pi_\rho(p)}\, .
\end{eqnarray}
The ``polarization functions" can be written as
\begin{equation}     \label{polarizationfunction}
\Pi_{\alpha} (p) = -i \int\frac{d^{4}q}{(2\pi)^{4}} tr[iS(q+p) \Gamma_{\alpha} i S(q)
\Gamma_{\alpha} ] \, ,
\end{equation}
where $\alpha = \sigma, \pi, \rho$ denotes the scalar, pseudoscalar and vector channel, respectively. The $\Gamma_{\alpha}$ is correspondingly $1$, $i\gamma_{5} \overrightarrow{\tau}$, and
$\gamma^{\mu} \overrightarrow{\tau}$ for the three channels.
It is remarkable that, even though the Eq.~(\ref{polarizationfunction}) is similar to that in the usual NJL model (see, {\it e.g.}, Refs.~\cite{Klevansky:1992Review,Huang:2015PRD}), it is in fact as the same as that in the Bethe-Salpeter equation when calculating the four quark interaction kernel~\cite{Wang:2013DSE,Fu:2013PRD,Roberts:2009PRL}.
To show this we take the terminology polarization function(s) with quotation marks in our context.

After some tedious calculations, one obtains the ``polarization functions" as
\begin{eqnarray}
&&\Pi_\sigma(p)=4iN_{c} N_{f} \big[ I_{1} -\frac{1}{2} (p^{2} - 4M^{2}) I(p) \big] \, , \\
&&\Pi_\pi(p)=4iN_{c} N_{f} \big[ I_{1} - \frac{1}{2}p^{2} I(p) \big]  \, , \label{Pipi} \\
&&\Pi_\rho(p)=-8i N_{c} N_{f} \big[ I_{1} - \frac{1}{2} (p^{2} + 2 M^{2}) I(p) \big]  \, ,
\end{eqnarray}
where $M$ is the constituent quark mass, and
\begin{eqnarray}
I_{1} &=&\int\frac{d^4q}{(2\pi)^4}\frac{1}{q^2-M^2}\, , \\
I(p)  &=&\int\frac{d^4q}{(2\pi)^4}\frac{1}{ [ (q+p)^2-M^2 ] (q^2-M^2)}\, . \label{I}
\end{eqnarray}
From the pole of the meson propagators, we can obtain the meson mass from equation
\begin{equation}
1-2g_\alpha\Pi_\alpha(m_\alpha)=0 \, . \label{pole}
\end{equation}

The pion decay constant $f_\pi$ can be calculated from the vacuum to
one-pion axial-vector matrix element. After some calculations we
have the following form for $f_{\pi}^{}$
\begin{equation}
f_{\pi}^{2} = -4 i N_{c}^{} M^2 I(0)\, , \label{fpi1}
\end{equation}
where $I(0)$ is defined in Eq.~(\ref{I}), but with $p=0$.

So far, we have only given formulas for the case of zero temperature.
To take into account the effect of finite temperature, we adopt in this paper the Matsubara formalism.
In this formalism, the energy part is replaced by the Matsubara frequencies $i\omega_{n}^{}$
with $\omega_{n}^{} = 2n{\pi}T$ for bosons and $\omega_{n}^{} = (2n+1){\pi}T$ for
fermions. Then the integral can be given as
\begin{equation}
\int\frac{d^4p}{(2\pi)^4}f(p^0,\overrightarrow{p}) = i\beta^{-1} \sum_n \int\frac{d^3p}{(2\pi)^3}f(i\omega_{n}^{},\overrightarrow{p}) \, ,
\end{equation}
where $\beta=1/T$ is the inverse of temperature. Then the gap equation can be rewritten by
\begin{equation}
M=m_{0}^{} + 4{g_{s}^{}} N_{c}^{} N_{f}^{} \int\frac{d^3q}{(2\pi)^3}\frac{M}{E_{q}^{}} [ 1 - 2n_{f}^{}(E_{q}^{}) ] \, ,
\end{equation}
with
\begin{equation}
E_{q}^{} = \sqrt{q^2 + M^2}\, ,
\end{equation}
\begin{equation}
n_{f}^{} (E_{q}^{}) = \frac{1}{e^{{\beta}E_{q}^{}} + 1} \, .
\end{equation}
For the ``polarization functions", the integrals are
\begin{equation}
I_{1} = -i\int\frac{d^3q}{(2\pi)^3}\frac{1}{2E_{q}^{}}(1-2 n_{f}^{}(E_{q}^{})) \, , \label{I11}
\end{equation}
\begin{eqnarray}
I(p)& = & i\int\frac{d^3q}{(2\pi)^3}\frac{1}{4 E_{q}^{} E_{q+p}^{}} \Big{[} [ n_{f}^{}(E_{q}^{}) - n_{f}^{}(E_{q+p}^{}) ] \notag \\
&& \;\;\;\; \times \big{(} \frac{1}{p_{0}^{} + E_{q}^{} - E_{q+p}^{}}
- \frac{1}{p_{0}^{} - E_{q}^{} + E_{q+p}^{}} \big{)} \notag \\
&& + [1 - n_{f}^{} (E_{q}^{}) - n_{f}^{} (E_{q+p}^{}) ] \notag \\
&& \;\;\;\; \times \big{(} \frac{1}{p_{0}^{} + E_{q}^{} + E_{q+p}^{}} - \frac{1}{p_{0}^{} - E_{q}^{} - E_{q+p}^{}} \big{)} \Big{]} \, . \qquad
\end{eqnarray}
For the simple case that the external three-momentum $\vec{p}=0$,
$I(p)$ has a more simple form
\begin{equation}
I(p) = i\!\! \int \!\! \frac{d^3q}{(2\pi)^3} [ 1 - 2 n_{f}^{} (E_{q}^{}) ] \Big{(} \frac{1}{p_{0}^{} + 2E_{q}^{}}
-\frac{1}{p_{0}^{} - 2E_{q}^{}} \Big{)} \, . \label{I0}
\end{equation}
With Eqs.~(\ref{I11}) and (\ref{I0}) we can solve Eq.~(\ref{pole}) to
obtain the mesons' masses. One can notice easily then
\begin{equation}
(m_{\sigma}^{2} - 4 M^{2} ) I(m_{\sigma}^{}) = m_{\pi}^{2} I (m_{\pi}^{})  \, . \label{GT-relationO}
\end{equation}
Since the function $I(p)$ is usually a very smooth function of $p$ and depends on $p$ quite weakly~\cite{Klevansky:1992Review}, one can have then approximation $I(m_{\sigma}^{}) = I(m_{\pi}^{}) = I(0)$. As a consequence, one can have
\begin{equation}
m_{\sigma}^{2}  = m_{\pi}^{2} + 4 M^{2}    \, . \label{GT-relation}
\end{equation}
%
%
It is apparent, as there exist  chiral symmetry exactly, $M = 0$, one has $m_{\sigma}^{} = m_{\pi}^{}$.
Therefore the degeneracy of the $\sigma$ meson and pion masses is usually regarded as a signal of the chiral symmetry restoration.

\section{In an external magnetic field}

With an external magnetic field, the NJL Lagrangian is given as
\begin{eqnarray}
\mathcal{L} & = & \overline{\psi}(i\partial \!\!\!/
- q_{f}^{} eA \!\!\!/ - m_{0}^{})\psi
+{g_{s}^{}} [(\overline{\psi}\psi)^2+(\overline{\psi}i\gamma_{5}^{} \overrightarrow{\tau}\psi)^2] \notag \\
&& -{g_{v}^{}} (\overline{\psi}\gamma^\mu \overrightarrow{\tau} \psi)^2\, ,
\end{eqnarray}
where $q_{f}^{}$ is the quark electric charge number, $2/3$ for up quark and $-1/3$ for down quark.
We assume a homogeneous external magnetic field $B$ along the $z$-direction,
then $A$ can be chosen as
\begin{equation}
A=(0,-\frac{1}{2}Bx_2,\frac{1}{2}Bx_1,0).
\end{equation}
The quark propagator has the form
\begin{equation}
S(q)=\frac{1}{q \!\!\!/ - q_{f}^{} eA \!\!\!/ - M}
=\frac{q \!\!\!/ - q_{f}^{} eA \!\!\!/ + M}{(q - q_{f}^{} eA)^{2} - M^{2} } \, .
\end{equation}
After some calculations, part of the denominator of the above equation can be expressed as
\begin{equation}
(q - q_{f}^{} eA)^2 = q_{0}^{2} - (q_{3}^{2} + 2n|q_{f}^{}|eB)  \, ,
\end{equation}
which means that, because of the existence of the external magnetic field,
the transverse part of the three momentum which is perpendicular to the $z$-direction
is quantized as discrete Landau levels.
Then the $\Phi$-derivable scheme and all the equations in the previous section can be easily extended to finite magnetic field.
We list the main ones in the following.

The gap equation is given as
\begin{equation}
M = m_{0}^{} + 2{g_{s}^{}} N_{c}^{} \sum_{f} \! \frac{|q_{f}^{}|eB}{2\pi}\! \sum_{n=0}^\infty \!
\alpha_{n}^{} \!\! \int \!\! \frac{dq_{3}^{}}{2\pi}\frac{M}{E_{q}^{}} [1 - 2 n_{f}^{} (E_{q}^{}) ] \, ,
\end{equation}
with
\begin{equation}   \label{Eq2-general}
E_{q}^{2} = q_{3}^{2} + 2n|q_{f}^{}| eB + M^{2} \, ,
\end{equation}
and $\alpha_{n}^{}$ is the spin degeneracy factor,
\begin{equation}
\alpha_{n}^{} =
\begin{cases}
1& {n=0}\\
2& \text{otherwise}
\end{cases}.
\end{equation}

The ``polarization function" of $\sigma$ meson in a magnetic field is given by
\begin{equation}
\Pi_{\sigma}(p) = 2 i N_{c}^{} \sum_{f} \frac{|q_{f}^{}|eB}{2\pi} \big{[} I_{1}^{} - \frac{1}{2}(p^{2} - 4M^{2})I(p) \big{]} \, ,
\end{equation}
with
\begin{equation}
I_{1}^{} = - i \sum_{n} \alpha_{n}^{} \int\frac{d{q_{3}^{}}}{2\pi}\frac{1}{2E_{q}^{}} [1-2n_{f}^{} (E_{q}^{}) ] \, , \label{I1emf}
\end{equation}
\begin{eqnarray}
I(p)=&& i \sum_{n} \alpha_{n}^{} \int\frac{d{q_{3}^{}}}{2\pi}\frac{1}{4 E_{q}^{} E_{q+p}^{}}
\Big{[} [n_{f}^{} (E_{q}^{}) - n_{f}^{}(E_{q+p}^{}) ] \notag \;\; \\
&& \;\; \times \big{(} \frac{1}{p_{0}^{} + E_{q}^{} - E_{q+p}^{}} -\frac{1}{p_{0}^{} - E_{q}^{} + E_{q+p}^{}} \big{)} \notag \\
&& + [ 1 - n_{f}^{} (E_{q}^{}) - n_{f}^{} (E_{q+p}^{}) ] \notag \\
&& \;\; \times \big{(} \frac{1}{p_{0}^{} + E_{q}^{} +E_{q+p}^{}} - \frac{1}{p_{0}^{} - E_{q}^{} - E_{q+p}^{}}
\big{)} \Big{]} \, . \label{Ipemf}
\end{eqnarray}
Considering zero three momentum $\vec{p}=0$, the ``polarization function" can
be simply written as
\begin{eqnarray}
&&\Pi_{\sigma}(p_{0}^{}) = \frac{M - m_{0}^{}} {2 {g_{s}^{}} M}
+ 2 N_{c} \sum_{f} \frac{|q_{f}^{}|eB}{2\pi}\frac{1}{2}(p_{0}^{2} - 4 M^2)\sum_{n} \alpha_{n}^{} \notag \\
&& \;\; \times\! \int \! \frac{d{q_{3}^{}}}{2\pi}\frac{1}{4E_{q}^{2}} [1\! - \! 2 n_{f}^{} (E_{q}^{}) ]
\big{(} \frac{1}{p_{0}^{} - 2E_{q}^{}} - \frac{1}{p_{0}^{} + 2 E_{q}^{}} \big{)} \, .
\end{eqnarray}

For the pion, we should notice that the form of the ``polarization function" in case of zero magnetic field includes the contributions of quark--antiquark loops for both $u$ ($\bar{u}$) and $d$ ($\bar{d}$) quarks, thus the mass we get is actually that of the neutral pion $\pi^{0}$. It is also obvious that without magnetic field we cannot distinguish the charged pion $\pi^{\pm}$ from the neutral $\pi^{0}$ in view of the ``polarization functions".
In fact, when determining the parameters, one usually fixes the pion mass as the $\pi^{0}$ mass. When considering finite magnetic field, everything changes. The isospin Pauli matrix for neutral $\pi^{0}$ takes $\tau^{3}$, and for charged $\pi^{\pm}$ takes $\tau^{\pm} =(\tau^{1}{\pm} i\tau^{2})/\sqrt{2}$. For the neutral $\pi^{0}$ we can directly get the ``polarization function", which has almost the same form as Eq.~(\ref{Pipi}) except the three momentum integrations being replaced by a sum of the Landau level and a integration of $p_{3}^{}$,
\begin{equation}
\Pi_{\pi^{0}}(p) = 2 i N_{c} \sum_{f} \frac{|q_{f}^{}|eB}{2\pi} \big{[} I_{1} -\frac{1}{2}p^{2} I(p)  \big{]} \, ,
\end{equation}
where $I_{1}$ and $I(p)$ are the same as Eq.~(\ref{I1emf}) and Eq.~(\ref{Ipemf}).
With $\vec{p}=0$,
\begin{eqnarray}
&&\Pi_{\pi^{0}}(p_{0}^{})=\frac{M - m_{0}^{}}{2{g_{s}^{}}M} + 2 N_{c} \sum_{f} \frac{|q_{f}^{}|eB}{2\pi}\frac{1}{2}p_{0}^{2} \sum_{n} \alpha_{n} \qquad\quad \notag \\
&&\times\! \int \! \frac{d{q_{3}^{}}}{2\pi}\frac{1}{4E_{q}^{2}} [1 \! - \! 2 n_{f}^{}(E_{q}^{}) ]
\Big{(} \frac{1}{p_{0}^{} \! - \! 2E_{q}^{}} - \frac{1}{p_{0}^{} \! + \! 2 E_{q}^{}} \Big{)} \, .
\end{eqnarray}
For the charged $\pi^{\pm}$, we can start from the inner structure of $\pi^{\pm}$.
Different from $\pi^{0}$, $\pi^{+}$ is composed of $u$  and $\overline{d}$ quark,
so that the quark loop structure of summing over the $u$ ($\bar{u}$) and $d$ ($\bar{d}$) quarks for $\pi^{0}$ should be replaced by one $u$ quark and one $\overline{d}$ quark for $\pi^{+}$.
Similarly, one can have that for $\pi^{-}$.
Thus the ``polarization function" has the form (here we also consider the case of zero three momentum)
\begin{equation}
\Pi_{\pi^{\pm}}(p) = 2 i N_{c} \frac{eB}{2\pi} \big{[} I_{1} - \frac{1}{2} p^{2} I(p) \big{]}  \, ,
\end{equation}
with
\begin{eqnarray}
I(p)=&&i\sum_{n} \alpha_{n} \int\frac{dq_3}{2\pi}\frac{1}{4E_{q} E_{q}^{\prime}}
\Big{[} [ n_{f}^{}(E_{q}^{}) - n_{f}^{}(E_{q}^{\prime}) ] \notag \\
&& \times \big{(} \frac{1}{p_{0}^{} + E_{q} - E_{q}^{\prime}}
- \frac{1}{p_{0}^{} - E_{q}^{} + E_{q}^{\prime}} \big{)} \notag \\
&& + [ 1 - n_{f}^{} (E_{q}^{}) - n_{f}^{} (E_{q}^{\prime}) ]  \notag \\
&& \times \big{(} \frac{1}{p_{0} + E_{q}^{} + E_{q}^{\prime}}
- \frac{1}{p_{0} - E_{q}^{} - E_{q}^{\prime}} \big{)} \Big{]} \, , \label{Ip2emf}
\end{eqnarray}
\begin{equation}
E_{q}^{2} = 2n|q_{u}^{}|eB + q_{3}^{2} + M^{2} \, ,
\end{equation}
\begin{equation}
{E_{q}^{\prime}}^{2} = 2 n^{\prime}|q_{d}^{}|eB + q_{3}^{2} + M^{2} \, .
\end{equation}
Due to the constraints of conservation of momentum,
$n$ and $n^{\prime}$ has relation $n^{\prime}=2n$.
Then Eq.~(\ref{Ip2emf}) can be reduced to a simple form
\begin{equation}
I(p)=i \sum_{n} \alpha_{n}^{} \!\! \int \!\! \frac{d{q_{3}^{}}}{2\pi} [ 1 - 2 n_{f}^{} (E_{q}^{}) ]
\Big{(} \frac{1}{p_{0}^{} \! + \! 2 E_{q}^{} } - \frac{1}{p_{0}^{} \! - \! 2 E_{q}^{} } \Big{)} \, ,
\end{equation}
with
\begin{equation}  \label{Eq2-chargedpion}
E_{q}^{2} =\frac{4}{3} n e B + q_{3}^{2} + M^{2}\, .
\end{equation}

Similar as pion, we can easily get the neutral $\rho^{0}$ and charged $\rho^{\pm}$ ``polarization functions"
\begin{eqnarray}
&&\Pi_{\rho^{0}}(p_{0}^{})=\frac{M-m_{0}^{}}{{g_{s}^{}}M} +4N_{c} \sum_{f} \frac{|q_{f}^{}|eB}{2\pi}\frac{1}{2}(p_{0}^{2}+2M^{2})\sum_{n}\alpha_{n}^{} \notag \\
&& \times\!\! \int \!\! \frac{d{q_{3}^{}}}{2\pi}\frac{1}{4E_{q}^{2}} [1-2 n_{f}^{}(E_{q}^{})]
\Big{(} \frac{1}{p_{0}^{} \! - \! 2E_{q}^{}} -\frac{1}{p_{0}^{}\! + \! 2E_{q}^{}} \Big{)} \, ,
\end{eqnarray}
with the $E_{q}^{}$ expressed in Eq.~(\ref{Eq2-general});
%

\begin{eqnarray}
&&\Pi_{\rho^{\pm}}(p_{0}^{})=\frac{M-m_{0}^{}}{{g_{s}^{}}M}+4N_{c} \frac{eB}{2\pi}\frac{1}{2}(p_{0}^{2} +2M^{2})\sum_{n}\alpha_{n}^{} \qquad \notag \\
&&\times \! \! \int \!\! \frac{d{q_{3}^{}}}{2\pi}\frac{1}{4E_{q}^{2}} [ 1 \! - \! 2 n_{f}^{} (E_{q}^{}) ]
\Big{(} \frac{1}{p_{0}^{} \! - \!  2 E_{q}^{}} - \frac{1}{p_{0}^{} \! + \! 2 E_{q}^{}} \Big{)} \,  ,
\end{eqnarray}
with the $E_{q}^{}$ expressed in Eq.~(\ref{Eq2-chargedpion}).
%
%

Together with Eq.~(\ref{pole}) we can get the corresponding meson mass.

The magnetic field strength dependence of the quark condensate at finite temperature can still be
determined by Eq.~(\ref{phi}) with the corresponding quark mass $M$. And
the pion decay constant can be given by
\begin{equation}
f_{\pi}^{2} = -i N_{c} M^{2} \sum_{f} \frac{|q_{f}^{}|eB}{2\pi} I(0), \label{fpi2}
\end{equation}
where $I(0)$ is determined by Eq.~(\ref{Ipemf}) with $p=0$.

\section{Numerical results and discussions}

The feature of four fermion contact interactions of the NJL model makes the model nonrenormalizable,
an effective three momentum cutoff $\Lambda$ is thus needed to regulate the divergent quantities.
Together with the small bare quark mass $m_{0}^{}$, scalar interaction constant $g_{s}^{}$,
vector interaction constant $g_{v}^{}$, there are four parameters in the NJL model.
The parameters $\Lambda$, $m_{0}^{}$ and $g_{s}^{}$ are usually taken as~\cite{Buballa:2003Review}  $\Lambda=587.9\;$MeV, $m_{0}^{}=5.6\;$MeV, ${g_{s}^{}}\Lambda^{2} = 2.44$ which are fixed by fitting the quantities at zero temperature: $f_{\pi}^{}=92.4\;$MeV, $m_{\pi}^{}=135.0\;$MeV and  $\langle\overline{u}u\rangle^{1/3}=-240.8\;$MeV.
The last parameter $g_{v}^{}$ is fixed as $g_{v}^{}=1.39\times10^{-6}\, \textrm{MeV}^{-2}$
by fitting the zero temperature $\rho$ meson mass $m_{\rho}^{} =770.0\;$MeV.

In case of strong magnetic field, the sharp three momentum cutoff
$\theta(\Lambda - |\vec{p}|)$ suffers from a cutoff artifact since the continuum momentum is replaced by the discrete Landau quantized one.
To avoid this problem, a smooth cutoff ${f_{\Lambda}^{}}(\vec{p})$ is
introduced~\cite{Fukushima:2010PRD} as
\begin{equation}
{f_{\Lambda}^{}}(\vec{p}) = \sqrt{\frac{\Lambda^{2N}}{\Lambda^{2N} + |\vec{p}|^{2N}}} \, .
\label{eqn:smooth-cutoff}
\end{equation}
It is apparent that, in the limit of $N \rightarrow \infty$, ${f_{\Lambda}^{}}(\vec{p})$ is reduced to the sharp cutoff form.
In our practical calculation, we take the Eq.~(\ref{eqn:smooth-cutoff}) with $N=10$ for the cutoff parameter and the commonly used values listed above for the other three parameters.

\begin{figure}[htp]
\centerline{\includegraphics[width=0.45\textwidth]{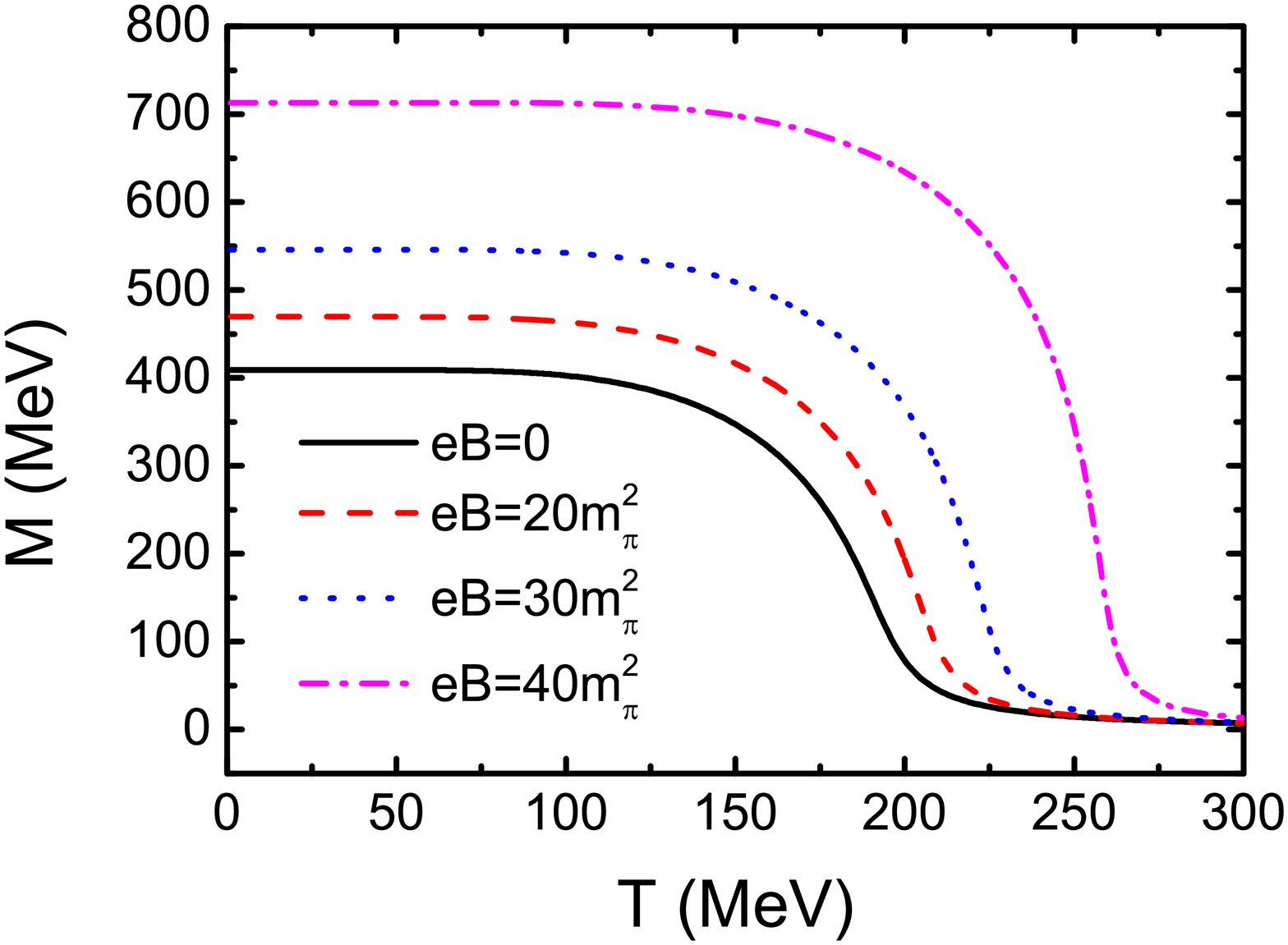}}
\caption{\label{fig:constituentquarkmass} (color online) Calculated variation behavior of the constituent quark mass
as a function of temperature at several values of magnetic field strength.
\emph{}  }\label{f1}
\end{figure}

\begin{figure}[htp]
\centerline{\includegraphics[width=0.45\textwidth]{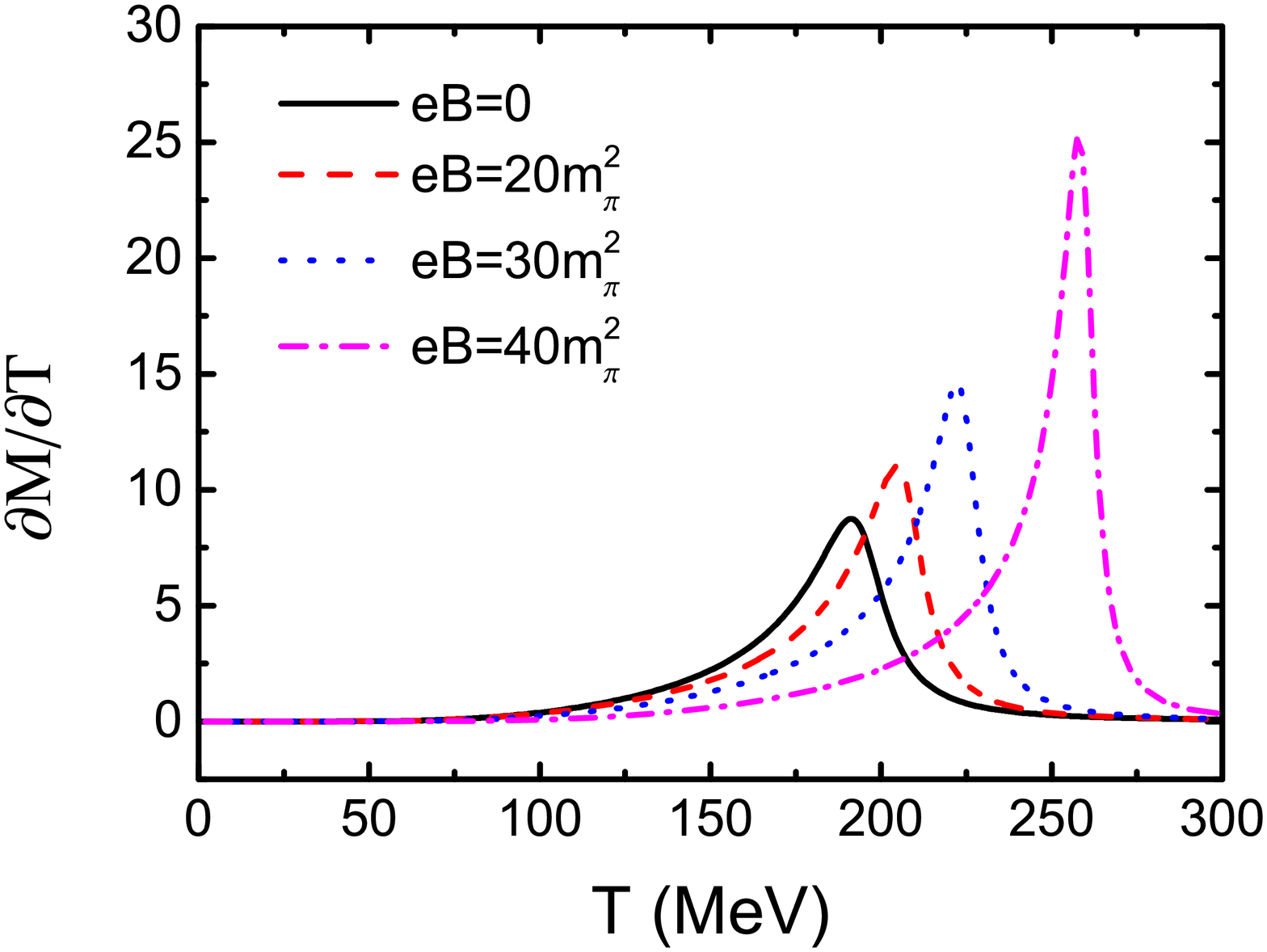}}
\caption{\label{fig:partialMpartialT} (color online) Calculated chiral susceptibility ${\partial}M/{\partial}T$ as a function
of $T$, corresponding to the same values of $eB$ in Fig.~\ref{f1}, respectively.
\emph{}  }\label{f2}
\end{figure}

In Fig.~\ref{f1} we plot the calculated variation behavior of the constituent quark mass as a function of
temperature with several values of magnetic field strength.
One can see that the constituent quark mass decreases quickly around certain temperature
for all the values of magnetic field strength.
We also plot the generalized chiral susceptibility ${\partial}M/{\partial}T$ as a function of temperature in Fig.~\ref{f2}.
From the position of the peak of ${\partial}M/{\partial}T$, we can obtain, as usual,  the
chiral phase transition temperature for different magnetic field strength,
{\it e.g.}, $T_{c}=191\;$MeV at $eB=0$.
When there is a finite magnetic field, Fig.~\ref{f1} shows that
the constituent quark mass increases with the magnetic field,
and the strength of the phase transition increases with the magnetic field strength as well,
as shown explicitly in Fig.~\ref{f2}, where both the height of the peaks and the (pseudo-)critical temperature increase with $eB$.

\begin{figure}[htb]
\centerline{\includegraphics[width=0.45\textwidth]{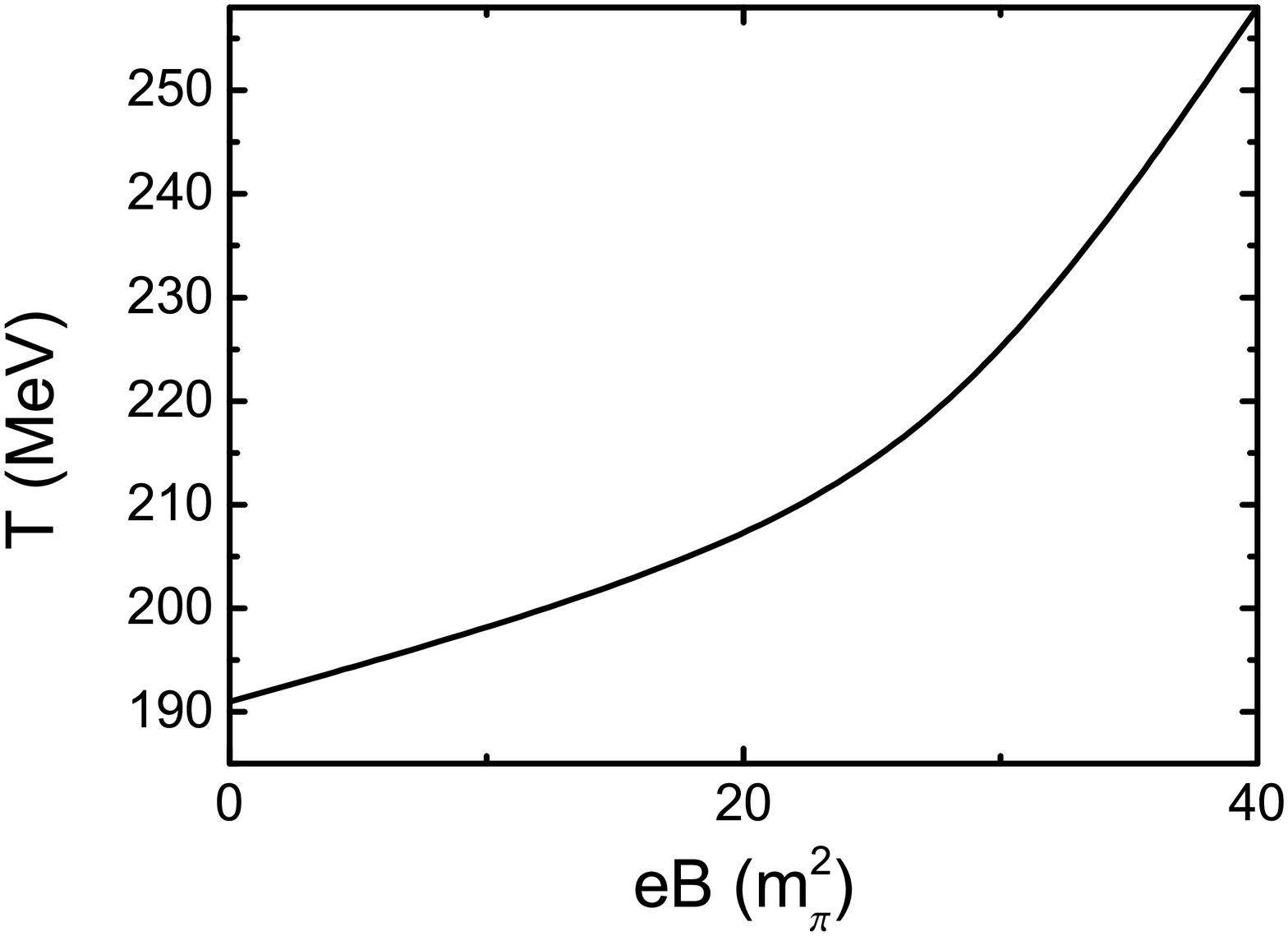}}
\caption{\label{fig:phasediagram}  Calculated phase diagram in the $T$--$eB$ plane.
\emph{}  }\label{f3}
\end{figure}

\begin{figure}[htp]
\centerline{\includegraphics[width=0.45\textwidth]{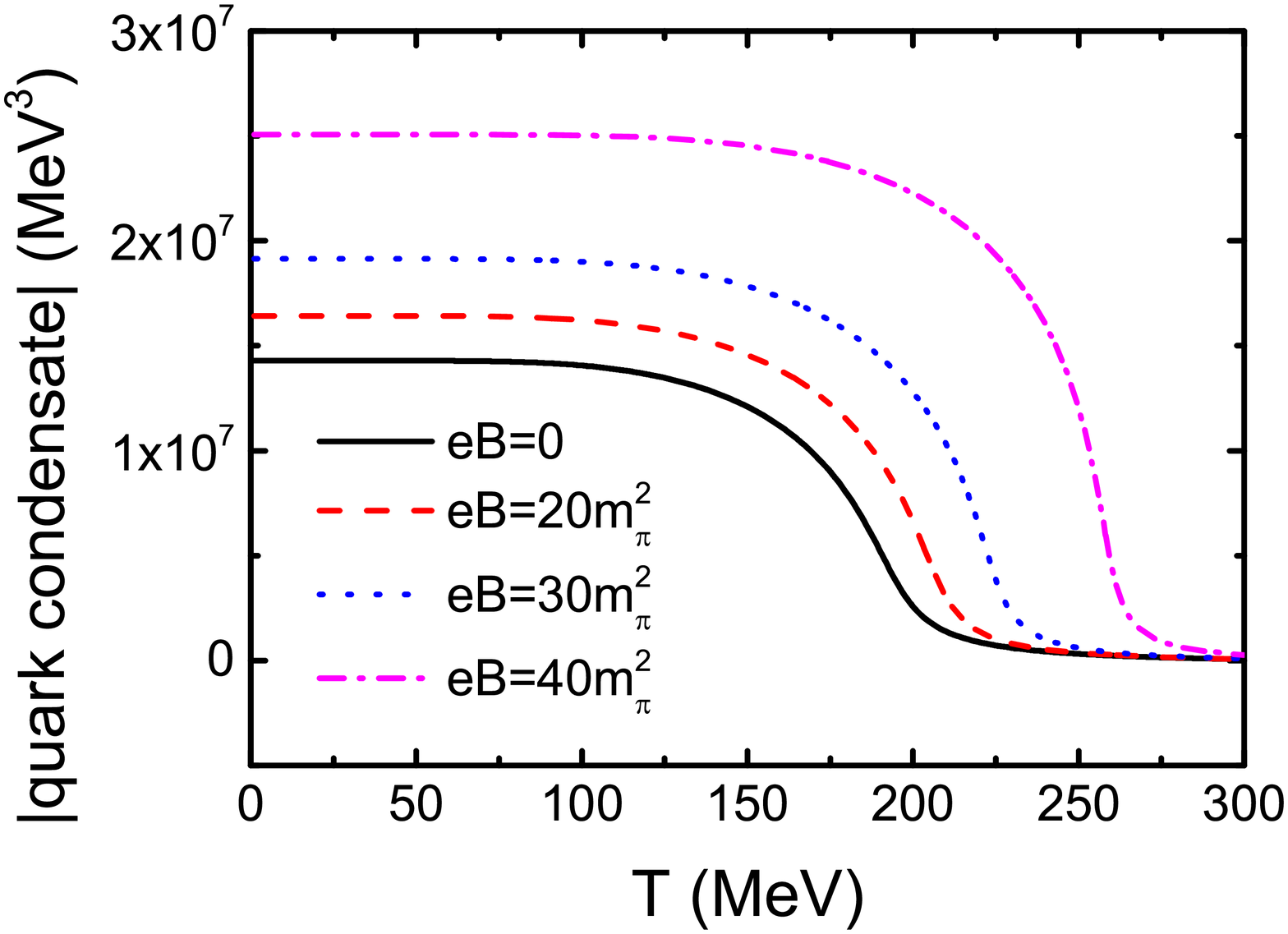}}
\caption{\label{fig:quarkcondensate}  (color online) Calculated absolute value of the quark condensate as a function of the temperature
at the same values of $eB$ as those in Fig.~\ref{f1}.
\emph{}  }\label{f4}
\end{figure}

We show the critical temperature, or the pseudo critical temperature more exactly,
as a function of $eB$ in Fig.~\ref{f3} and list some of the values in Table.~\ref{Tab:Critical-T-glance1}.
One can find from the figure and the Table that the critical temperature increases with $eB$, in particular for large magnetic field.
This result implies a confirmation of the magnetic catalysis in the $\Phi$--derivable scheme with the NJL model.
Fig.~\ref{f3} can also be treated as a phase diagram in the $T$--$eB$ plane,
where the regions above and below the curve correspond to the chiral symmetric phase and
the DCSB phase, respectively.
We also plot the absolute value of the quark condensate as a function of temperature in Fig.~\ref{f4}.
Because of Eq.~(\ref{phi}), we can get the same phase diagram as Fig.~\ref{f3}
via the criterion of $\partial\phi/{\partial}T$.
In addition, the phase boundary in the $T$--$eB$ plane can be parameterized as
\begin{equation}
 T_{c} = 191 + 1.827 (eB) - 0.109 (eB)^{2} + 0.00264 (eB)^{3} \, ,
 \end{equation}
with $eB$ in unit $m_{\pi}^{2}$.

\begin{table}[htb]
\caption{Calculated pseudo-critical temperature $T_{c}^{}$ 's in case without and with external magnetic field, obtained with different criteria and the melting temperature of $\rho$ meson, where $T_{c}^{\chi} $'s stand for those with the constituent quark mass $M$ and the chiral quark condensate
$\langle \bar{q} q \rangle$,  $T_{c}^{{\pi}^{0}}$ for the $\pi^{0}$ and $\sigma$ meson masses to begin to degenerate or nearly degenerate, $T_{c}^{f_{\pi}^{}} $ for the maximal decreasing rate of the $f_{\pi}^{}$,   $ T_{c}^{r}$ for the first minimum of the $r$, $T_{m}^{{\rho}^{0}}$ for the  $m_{{\rho}^{0}}^{}$ to degenerate with the $2M$, and $T_{m}^{{\rho}^{\pm}}$ the highest for the mass solution to  exist  (All the temperatures are in unit MeV and the $eB$ in $m_{\pi}^{2}$ at zero temperature and zero magnetic filed). } \label{Tab:Critical-T-glance1}
 \begin{tabular}
 {c|cccccc} \hline \hline $eB$  & ~~$T_{c}^{\chi} $~~ & ~~$T_{c}^{{\pi}^{0}}$~~ & ~~$T_{c}^{f_{\pi}^{}} $~~ &   ~~$ T_{c}^{r}$~~ &  ~~$ T_m^{{\rho}^{0}} $~~ &  ~~$ T_m^{{\rho}^{\pm}} $~~ \\  \hline
 0.0   & $191$ &  $233$  &  $191$  &  $200$  &  $155$   &   \\
 10.0 &  $201$ &  $256$  &  $201$  &  $211$  &  $194$  &  $169$ \\
 20.0 &  $205$ &  $277$  &  $205$  &  $214$  &  $215$  &  $173$ \\
 30.0 &  $222$ &  $304$  &  $222$  &  $232$  &  $245$  &  $196$ \\
 40.0 &  $258$ &    &  $258$  &    &  &   \\
\hline \hline
\end{tabular}
\label{Tc-glance}
\end{table}

\begin{figure}[htp]
\centerline{\includegraphics[width=0.40\textwidth]{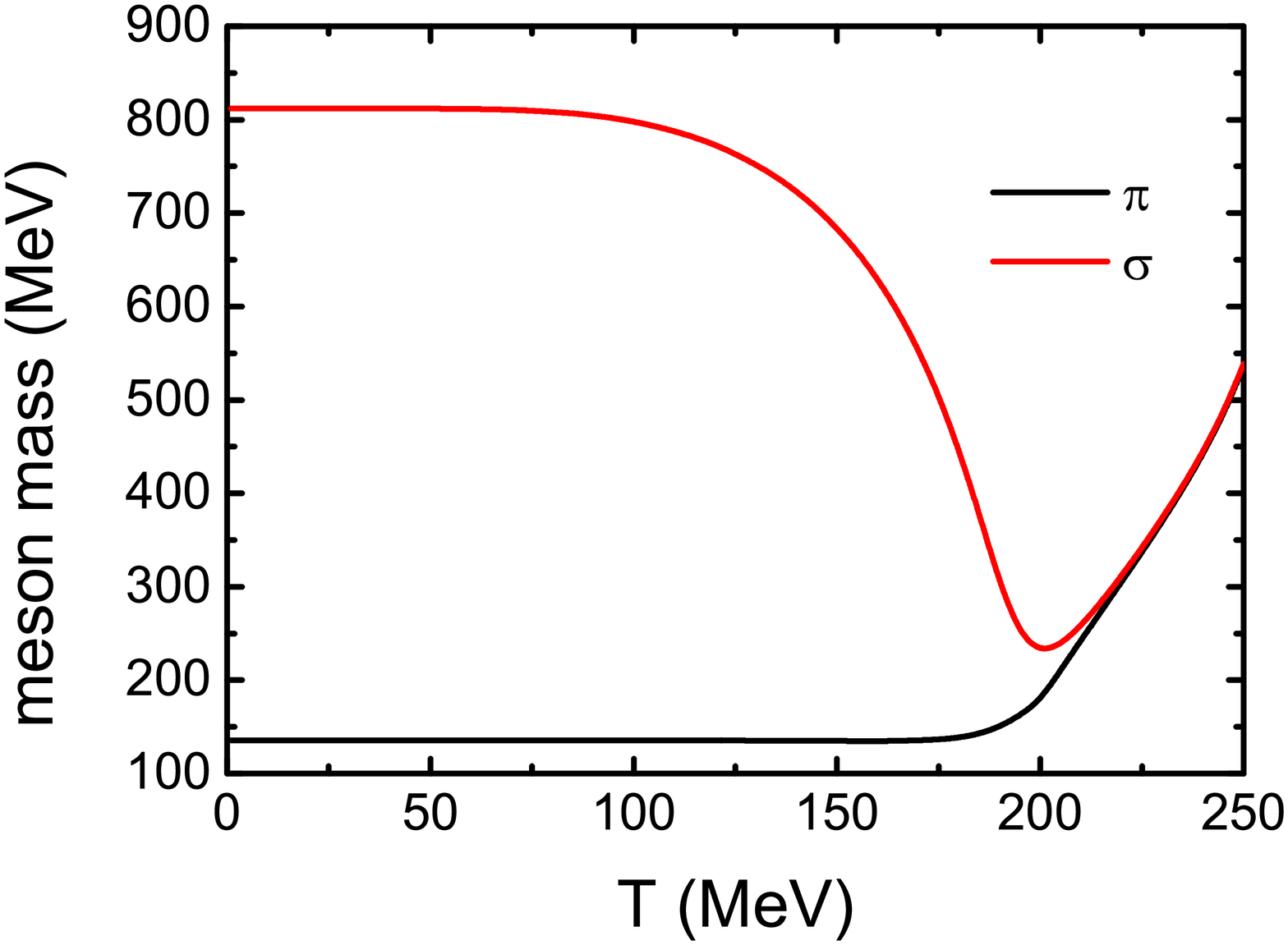}}
\caption{\label{fig:pisigmamasswithoutmagneticfield}
(color online) Calculated $\pi$ and $\sigma$ meson masses as functions of temperature $T$
in case of vanishing magnetic field.
\emph{}  }\label{f5}
\end{figure}

The calculated $\pi$ and $\sigma$ meson masses as functions of temperature
in case of zero magnetic field are plotted in Fig.~\ref{f5}.
Considering together with the variation behavior of the constituent quark mass,
one can notice that the relation
$m_{\sigma}^{2} = 4 M^{2} + m_{\pi}^{2}$ conserves precisely.
The degeneracy of the $\pi$ and $\sigma$ meson masses at high temperature implies evidently the restoration of the chiral symmetry.

Now we focus on the $\pi$ mass in case of finite magnetic field. For the charged $\pi^\pm$,
we need to consider the contribution not only from the internal constituent quark and antiquark,
but also from the point particle correction.
In Eq.~(\ref{mpi}), the point particle correction is given for the case of zero temperature.
We can directly extend it to the finite temperature case as
\begin{equation}
M^{2}_{\pi^{\pm}}(T,eB) = m^{2}_{\pi^{\pm}}(T,eB) + eB \, .
\end{equation}
In the above expression, $m_{\pi^{\pm}}(T,eB)$ denotes the pion mass calculated from the constituent quark and antiquark contribution, where we can make the simplification of zero external momentum.
This means that we consider the internal contribution to the meson mass in a static meson coordinate system.
When considering the pion as a point particle moving in the external magnetic field, the momentum of the pion perpendicular to the direction of the magnetic field is quantized as Landau level, and the lowest Landau level governs the ground state of the pion mass, so that the point particle correction is a kinetic effect.
The calculated masses of $\pi^{0}$ and $\pi^{\pm}$ in case of very weak magnetic field and zero temperature
are $135.1$~MeV, $142.4$~MeV, respectively.
Comparing with the experiment data $m_{\pi^{+}} = 139.6$~MeV, one can know that
the theoretical result of the $\pi^\pm$ mass agrees with experiment very well
(the error is only about $2\%$).

\begin{figure}[htp]
\centering
\includegraphics[width=0.45\textwidth]{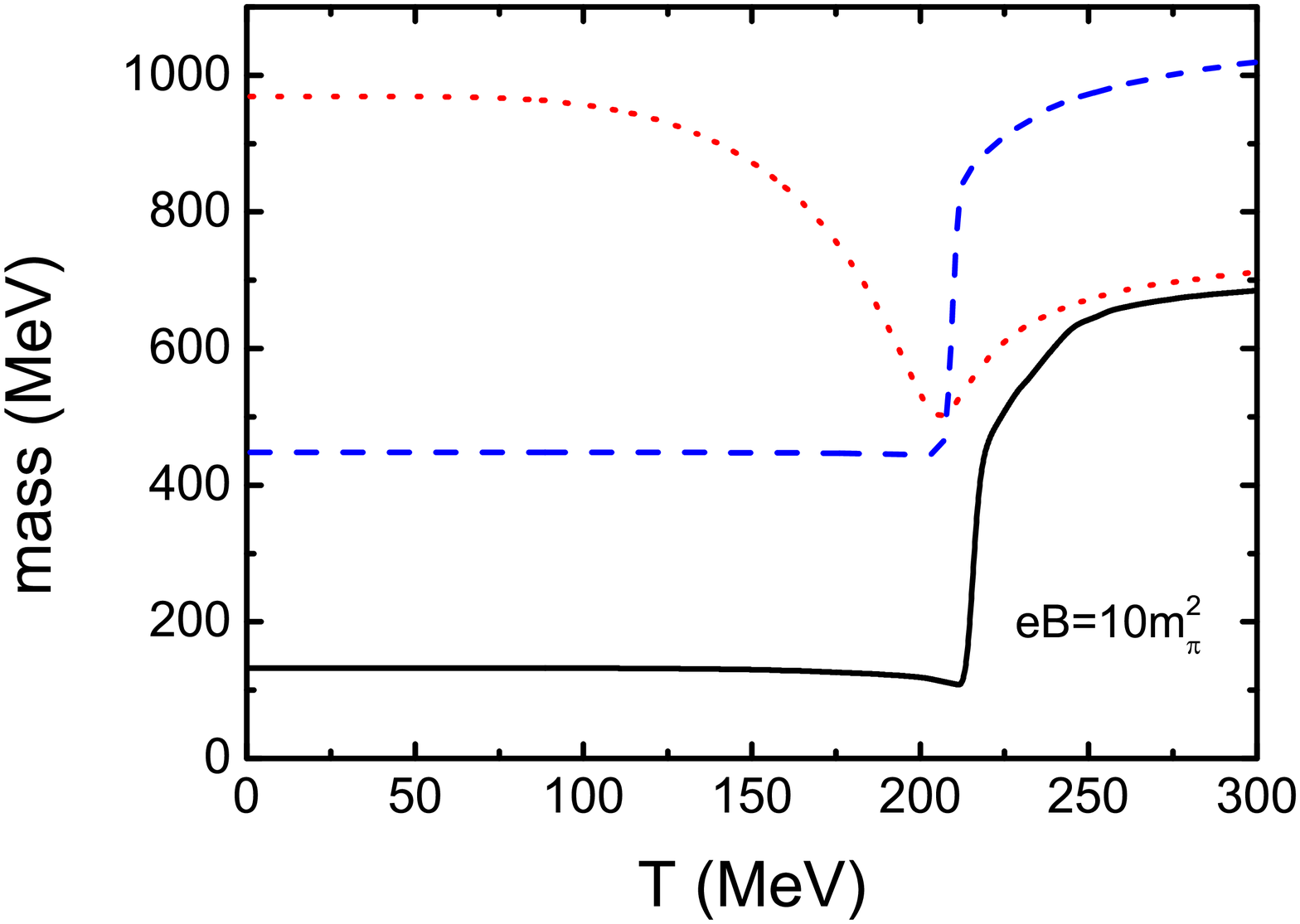}
\includegraphics[width=0.45\textwidth]{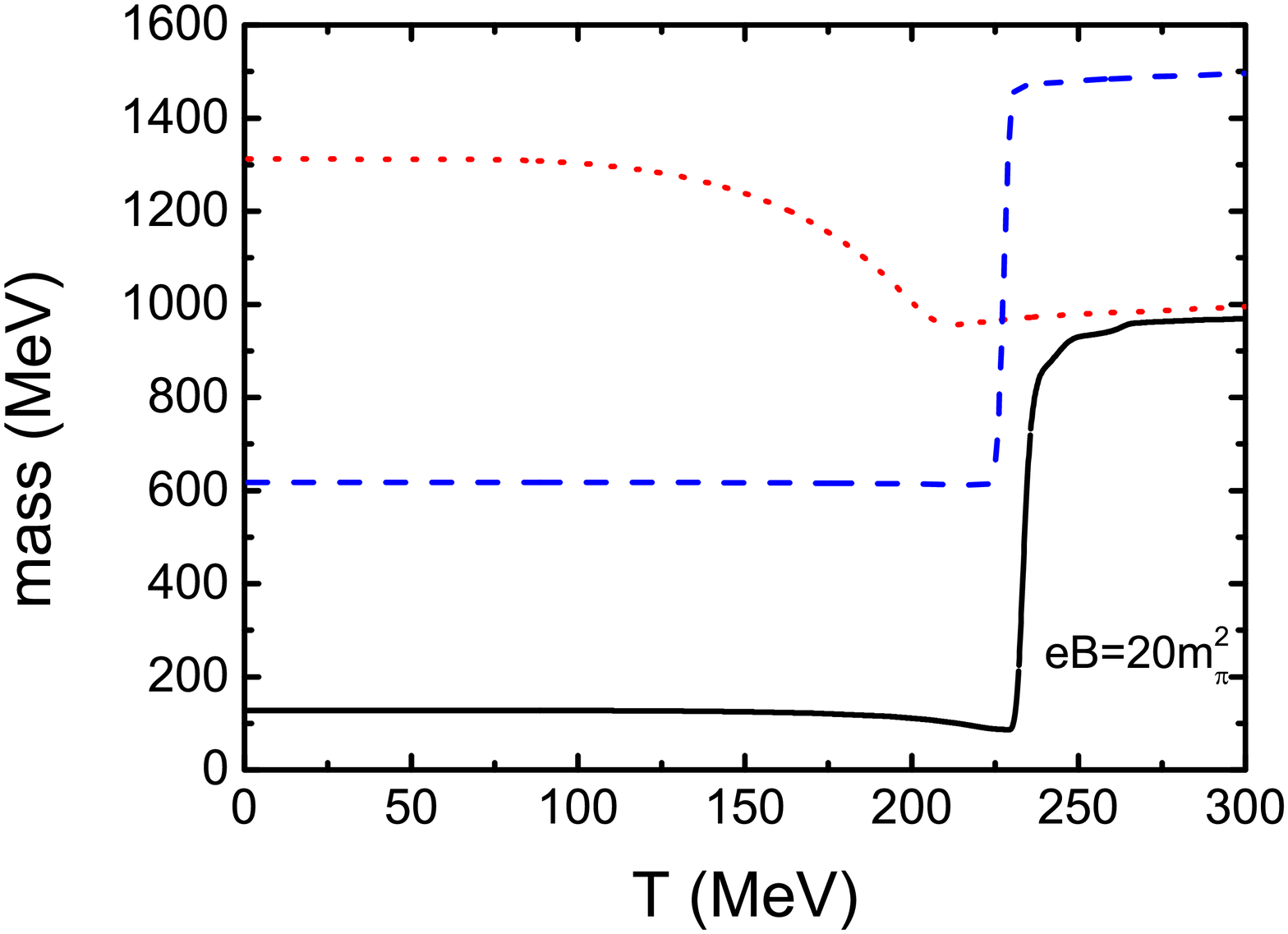}
\caption{(color online) Calculated $\pi^0$ and $\pi^\pm$ masses (in black solid, blue dashed line, respectively) together with $\sigma$ meson mass (in red dotted) as functions of temperature at two values of magnetic field strength.}\label{f6}
\end{figure}

In Fig.~\ref{f6} we illustrate the calculated masses of $\pi^{0}$ and $\pi^{\pm}$ mesons together with that of $\sigma$ meson as functions of temperature in two cases of nonzero magnetic field strength.
The $\sigma$ meson mass, at a fixed temperature, shows a monotonic increasing behavior with the  magnetic field strength. We can also find that in weak magnetic field the $\sigma$ meson mass keeps the same behavior as that in case of zero magnetic field, but with increasing the magnetic field strength the temperature dependence of the $\sigma$ meson mass becomes weaker, especially at the temperature around the (pseudo-)critical one. This feature indicates that the $\sigma$ meson mass depends on the magnetic field more drastically than on the temperature.
For the pions, it shows that there is almost no qualitative difference between the dependence of $\pi^{0}$ and $\pi^{\pm}$ masses on the temperature. When the temperature is lower than $191$~MeV, which is the (pseudo-)critical temperature of the chiral phase transition without magnetic field, $T_{c}^{\chi}$, the $\pi^{0}$ mass is almost a constant. Once the temperature gets higher than the critical value, the behavior becomes a little complicated. The $\pi^{0}$ mass decreases slightly around the $T_{c}^{\chi}$, and then increases suddenly so as to become nearly degenerate with the $\sigma$ mass when temperature is higher than the critical value $T_{c}^{\chi}$. Different from the zero magnetic field case, the degeneration is not so precise. The temperatures for the mass difference ($m_{\pi^{0}} - m_{\sigma}$) to be about $2\%$ of the $m_{\pi^{0}}$ are listed in Table~\ref{Tab:Critical-T-glance1}. The reason for the degeneracy to be not exact is the following. The existence of the magnetic field enhances the quark condensate and the constituent quark mass, but the temperature makes them to decrease and the constituent quark mass will drop to that in dynamical chiral symmetry at higher temperature. From the  relation between the masses of the $\sigma$-meson and the pion in Eq.~(\ref{GT-relation}), it is obvious that the degeneracy occurs at higher temperature. Another aspect is the finite current quark mass. When temperature is above the critical one, quark mass returns, in fact, to the current quark mass but not zero. Therefore the pion mass does not equate to the $\sigma$ mass precisely.
From Fig.~\ref{f6} we can also find that the critical temperature extracted from the ${\pi}^{0}$ mass increases with the magnetic field strength, which gives us the similar information of the phase diagram in $T$--$eB$ plane as shown in Fig.~\ref{f3}.
However, there is a difference between the critical temperatures in the two cases, which implies that the phase transition is not a sharp (low order) phase transition, but a crossover.
Moreover the $\pi^{\pm}$ mass increases with the magnetic field, no matter the temperature is lower or higher than the critical value. The critical temperature for $\pi^{\pm}$ mass to increase abruptly is almost the same as that for $\pi^{0}$ meson (only about $6\;$MeV lower).

\begin{figure}[htb]
\centerline{\includegraphics[width=0.45\textwidth]{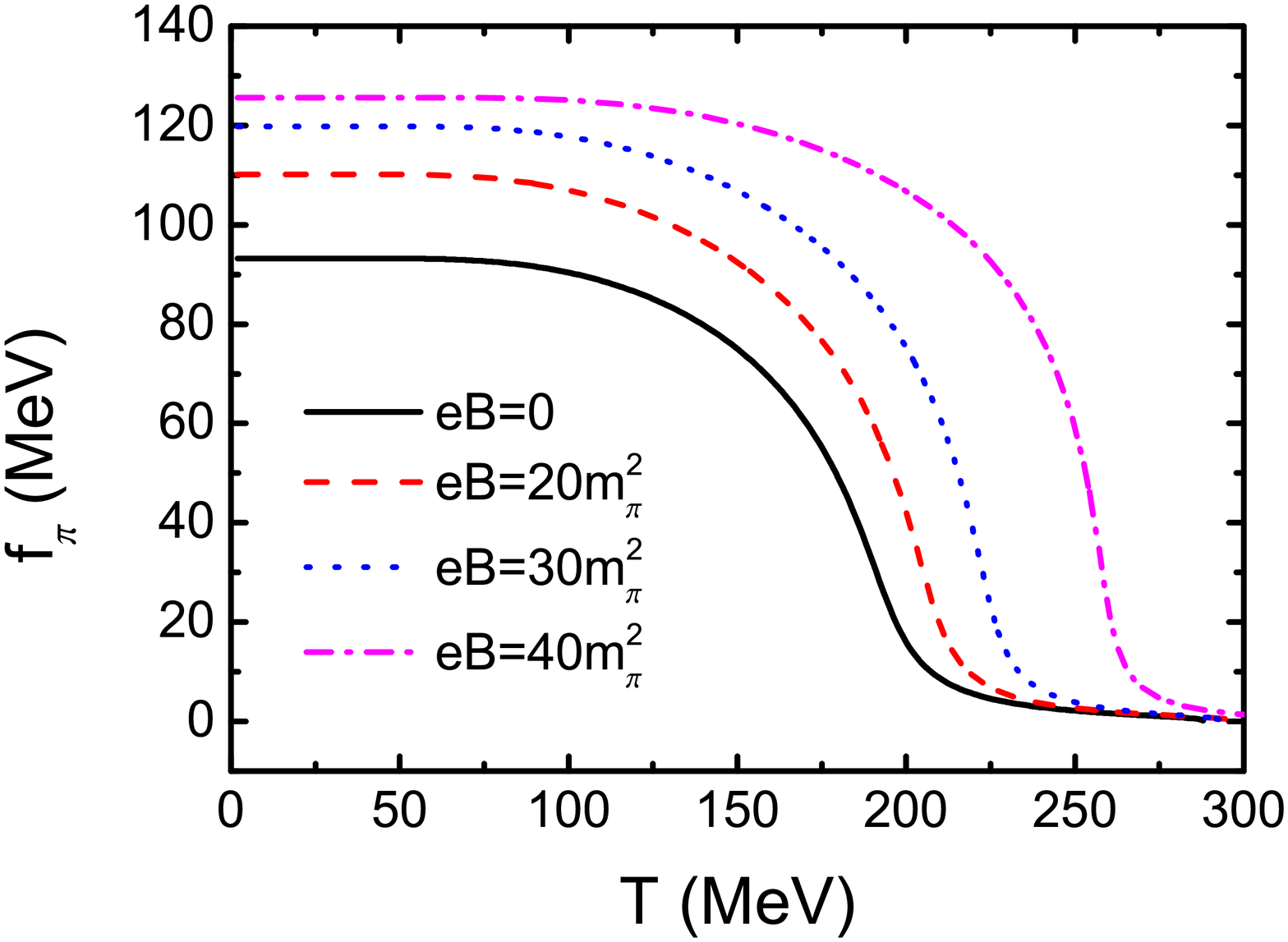}}
\caption{\label{fig:fpi} (color online) Calculated pion decay constant $f_{\pi}^{}$ as a function of  temperature at several values of magnetic field strength.
\emph{}  }\label{f7}
\end{figure}

It is well known that the Gell-Mann--Oakes--Renner (GOR) relation, which connects the $\pi$ mass and decay constant with the current quark mass and quark condensate, is a direct demonstration of the DCSB. The GOR relation reads
\begin{equation}
f_{\pi}^{2} m_{\pi}^{2} = - 2 m_{0}^{} \langle\overline{q}q\rangle \, ,
\end{equation}
where $2 \langle\overline{q}q\rangle$ includes the contributions of both the $u$ and $d$ quarks. From Eqs.~(\ref{fpi1}) and (\ref{fpi2}) we can get the $\pi$ decay constant. The obtained results at several values of the magnetic field strength are  displayed in Fig.~\ref{f7}. We should note that the $\pi$ decay constant in Eq.~(\ref{fpi1}) is related to the neutral pion $\pi^0$, so we only consider the decay constant for $\pi^0$ as shown in Eq.~(\ref{fpi2}), even though the $\pi^0$ and $\pi^{\pm}$ can be distinguished in a magnetic field. From Fig.~\ref{f7} we can notice that $f_{\pi}^{}$ at a certain temperature increases with the magnetic field strength; and at a fixed magnetic field, $f_{\pi}^{}$ decreases monotonously  with the increasing of temperature and falls to zero at high temperature, which is just qualitatively the same as that given in Ref.~\cite{Fayazbakhsh:2013PRD}. Comparing the variation behavior with those of the constituent quark mass and the quark condensate, one can find that the critical temperature at which the decreasing rate of the $f_{\pi}^{}$ takes its maximal value is exactly the same as that given with the constituent quark mass criterion (see Table I). It is easy to check that the GOR relation preserves very well at zero temperature and vanishing magnetic field. To examine the relation in case of finite magnetic field, following Ref.~\cite{Fu:2009PRD} we define a ratio
\begin{equation}
r=\frac{f_\pi^2m_\pi^2}{- 2 m_0\langle\overline{q}q\rangle}.\label{ratio}
\end{equation}

\begin{figure}[htb]
\centerline{\includegraphics[width=0.40\textwidth]{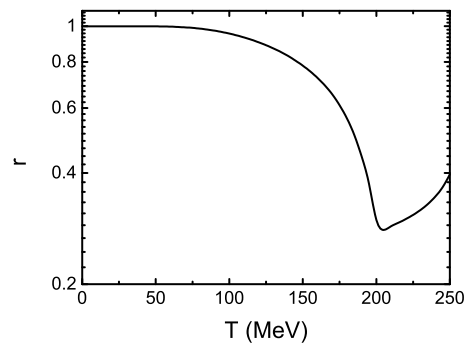}}
\caption{\label{fig:gorwithoutmagneticfield}
Calculated temperature dependence of the ratio $r$ defined in Eq.(\ref{ratio})
in case of vanishing magnetic field.
\emph{}  }\label{f8}
\end{figure}

\begin{figure}[htb]
\centerline{\includegraphics[width=0.45\textwidth]{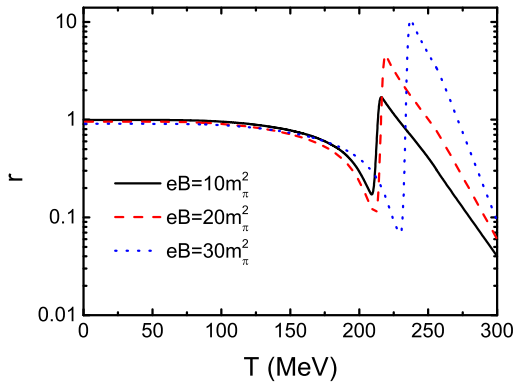}}
\caption{\label{fig:gor2} (color online)  Calculated temperature dependence of the ratio $r$ defined in Eq.(\ref{ratio}) at several values of magnetic field strength $eB$.
\emph{}  }\label{f9}
\end{figure}

We show the calculated result of the ratio as a function of temperature without magnetic field in Fig.~\ref{f8}. It shows obviously that, at low temperature, the ratio keeps almost a constant 1, which is a demonstration of the DCSB represented by the GOR relation. However, when  temperature increases, it deviates significantly from 1, which means that temperature damages the GOR relation drastically, or in other word, induces the dynamical chiral symmetry to be restored. Furthermore, we illustrate the dependence of the ratio on the temperature in case of nonzero magnetic field strength in Fig.~\ref{f9}. One can recognize easily from Fig.~\ref{f9} that with different strengths of the magnetic field, the ratio does not deviate from 1 distinctly either, if the temperature is lower than the critical one.  Once the temperature reaches up to around the critical value, $r$ fluctuates seriously and both the temperature for the fluctuation to reach its first minimum and that for it to take its maximum increase with the ascension of the magnetic field strength, which implies that the fluctuation of the ratio $r$ may be a signal for the chiral phase transition.  The temperatures for the fluctuation to take its first minimum in several cases of the magnetic field are listed in Table.~\ref{Tab:Critical-T-glance1}. These characteristics indicate that the external magnetic field preserves the DCSB.

Same as pions, we can get the masses of neutral and charged vector meson $\rho^0$ and $\rho^\pm$ from the vector ``polarization function" $\Pi_{\rho^0}(p)$ and $\Pi_{\rho^\pm}(p)$. For $\rho^\pm$, we also make the point particle correction,
\begin{equation}
M^2_{\rho^\pm}(T,eB)=m^2_{\rho^\pm}(T,eB)-eB.
\end{equation}

\begin{figure}[htb]
\centerline{\includegraphics[width=0.45\textwidth]{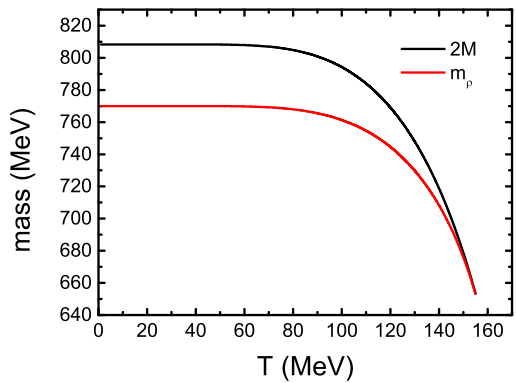}}
\caption{\label{fig:mrho} (color online)
Calculated $\rho$ meson mass and twice the constituent quark mass as functions of temperature in  case of vanishing magnetic field.
\emph{}  }\label{f10}
\end{figure}

\begin{figure}[htb]
\centerline{\includegraphics[width=0.45\textwidth]{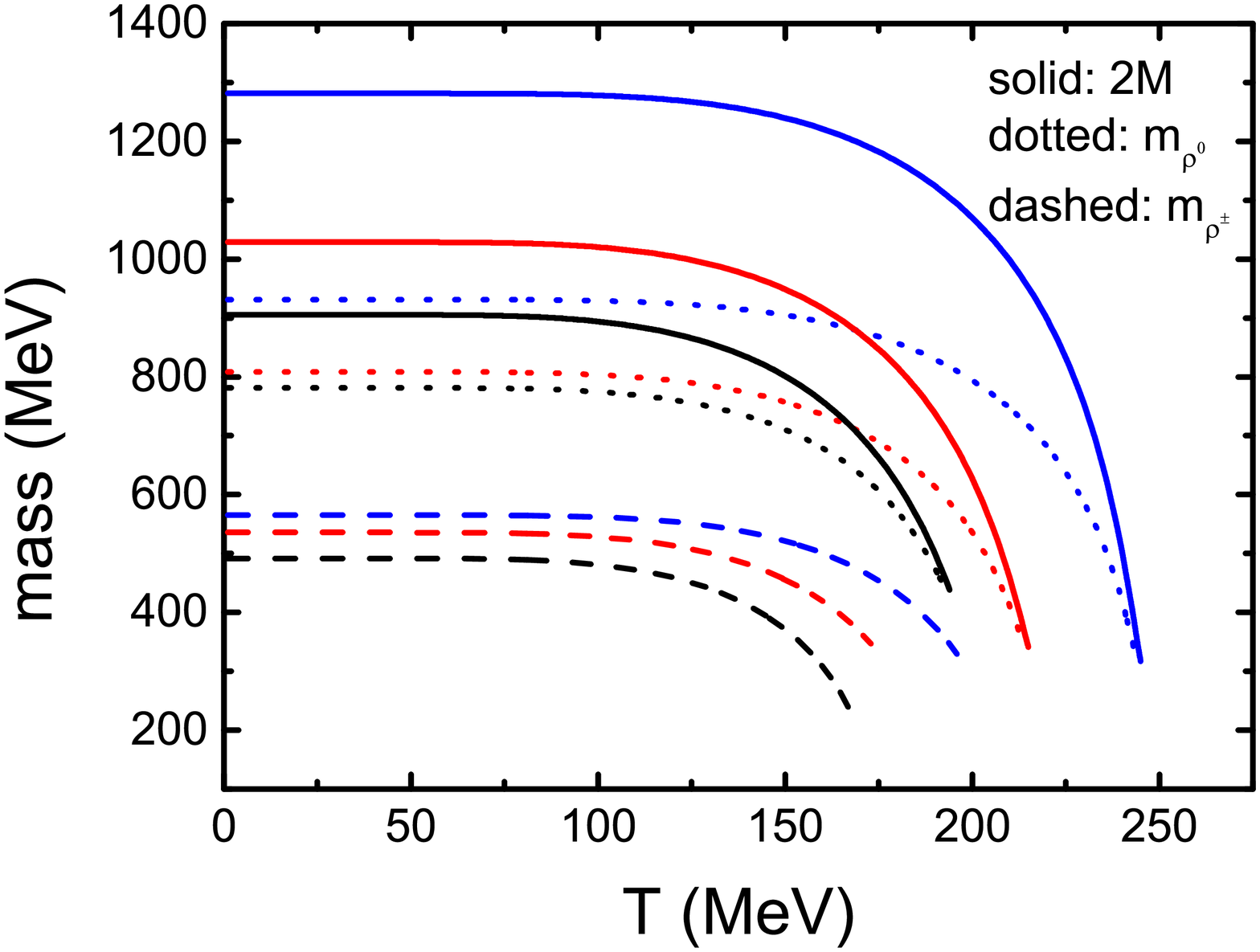}}
\caption{\label{fig:mrho2} (color online) Calculated $\rho$ meson masses and twice the constituent quark mass as functions of temperature at several values of magnetic field strength. The black lines stand for the results with $eB = 10 m_{\pi}^{2}$,  red for those with $eB=20 m_{\pi}^{2}$, and blue for $eB = 30 m_{\pi}^{2}$.
\emph{}  }\label{f11}
\end{figure}

We consider at first the case of vanishing magnetic field where we can not distinguish the charged $\rho^\pm$ from the neutral $\rho^{0}$, and illustrate the temperature dependence of the $\rho$ meson mass on the temperature in Fig.~\ref{f10}.
The figure displays evidently that $\rho$ meson mass decreases with temperature. When $T=155\;$MeV, the $\rho$ meson mass falls to the value of twice the constituent quark mass and there is no longer solution for the  $\rho$ meson mass at higher temperature.
This phenomenon implies that at a critical temperature the $\rho$ meson gets disassociated, or melt to two quarks (more exactly, a quark and an antiquark), and in turn, there is no $\rho$ meson condensate at high temperature.
In Fig.~\ref{f11} we plot the results of both $\rho^{0}$ and $\rho^{\pm}$ meson masses and twice the constituent quark mass as functions of temperature in case of nonzero magnetic field strength.
It shows that with finite magnetic field the $\rho^{0}$ meson mass has the same behavior as that in case without magnetic field, i.e., the $\rho^{0}$ meson will melt at the critical temperature when its mass is equal to the twice of the constituent quark mass and the melting temperature increases with magnetic field strength.
For the charged $\rho^{\pm}$ in a magnetic field, the mass also decreases with temperature and maintains smaller than the mass of $\rho^{0}$ meson.
Similar with the behavior of $\rho^0$, there are no longer solutions for the  $\rho^{\pm}$ mesons
as the temperature gets higher than a critical value.
It indicates that the $\rho^{\pm}$ mesons also melt at high temperature, and the melting temperature is lower than that for the $\rho^{0}$ meson in the same magnetic field.
All the melting temperatures of $\rho^{0}$ and $\rho^{\pm}$ are also listed in Table~\ref{Tab:Critical-T-glance1} for comparison.

Looking over Table~\ref{Tab:Critical-T-glance1}, one can recognize that  not only the pseudo-critical chiral symmetry restoration temperatures determined with different criteria but also the $\rho$ meson melting temperatures in case without magnetic field are all smaller than those in nonzero magnetic field strength. And the temperatures increase with strengthening the magnetic field. These features indicate that the external magnetic field can at least maintain the DCSB, so that there may exist magnetic catalysis in the region of the magnetic field strength we have considered.
Comparing the melting temperatures of the neutral $\rho$ meson and the charged $\rho$ meson with the pseudo-critical temperature of the chiral phase crossover, one can find that the $\rho$ mesons will get melted as the DCSB is still quite strong if the magnetic field is not strong enough
(for instance, $eB < 30 m_{\pi}^{2}$ for $\rho^{\pm}$ and $eB < 10 m_{\pi}^{2}$ for $\rho^{0}$), and in turn, there may not have vector meson condensates in the QCD vacuum, which is consistent with the lattice QCD result~\cite{Hidaka:2013lattice} and the model calculation results~\cite{Andreichikov:2013Hamiltonian,Taya:2015PRD}. One may also infer that there exists magnetic inhibition for vector hadrons.

\section{An Alternative View}

Considering the structure of the $\sigma$ meson discussed above, one can realize that it is the one having the quantum numbers of the vacuum, so that it plays a significant role in labeling the dynamical chiral symmetry restoration. However it most likely does not correspond to the meson observed in QCD~\cite{Chen:2007PRD,Fischer:2012PLB,Ananthanarayan:2001PR,Bugg:2004PR,Nakamura:2010Review,Oller:2012PRD}, since it has been well known that the $\sigma$ meson and the $\rho$ meson could be recognized as the resonant states of the $\pi$--$\pi$ scattering (see, {\it e.g.},  Refs.~\cite{Ananthanarayan:2001PR,Bugg:2004PR,Nakamura:2010Review,Oller:2012PRD}).
In order to check the results we obtained in last section, we re-calculate the temperature and magnetic field strength dependence of the masses of the $\sigma$ meson and $\rho$ meson
in the Roy equation~\cite{Roy:1971PLB} formalism of $\pi$--$\pi$ scattering~\cite{Fu:2009PRD,Schulze:1995JPG,He:1998NPA,Ananthanarayan:2001PR,Colangelo:2001NPB,Surovtsev:2008NPA,Caprini:2006PRL,Pelaez:2011PRL}. To analyze the stability of the mesons in magnetic field,
we also calculate the widths of the mesons' mass poles.

It has been well known that the significant inputs to determine the masses and their widths of the resonant states in $\pi$--$\pi$ scattering in the Roy equation scheme are the $\pi$--$\pi$ scattering lengths~\cite{Roy:1971PLB,He:1998NPA,Ananthanarayan:2001PR,Colangelo:2001NPB,Surovtsev:2008NPA,Caprini:2006PRL,Pelaez:2011PRL},  which can be determined by the mass and the decay constant of the pion and the relation in case of vanishing temperature and magnetic field has been well described in Ref.~\cite{Schulze:1995JPG}.
For our convenience we outline the scheme and quote only the main formulaes as the follows.
For the channel with isospin $I$, the scattering length $a_{I}^{}$ can be determined by the scattering amplitudes $T_{i}$ ($i=a,c,d,e$ stands for the mode of scattering represented in terms of the Feynman diagrams shown in the figure~1 of Ref.~\cite{Schulze:1995JPG} ), which can be fixed by the pion--quark-quark coupling constant, the ``polarization functions" and so forth.
After some calculation one can have~\cite{Schulze:1995JPG}
\begin{eqnarray}
a_{0}^{} &\! = \! & \frac{7}{32\pi} \Big{(} \frac{m_{\pi}^{}}{f_{\pi}^{}} \Big{)}^{2}  \big{[} 1 + \mathcal{O}( m_{\pi}^{2} ) \big{]} \, ,   \\
a_{1}^{} &\! = \! & 0 \, , \\
a_{2}^{} &\! = \! & \frac{-2}{32\pi} \Big{(}\! \frac{m_{\pi}^{}}{f_{\pi}^{}} \! \Big{)}^{2}
\Big{[} 1 \! - \! \big{(} 1 \! - \! 5 z \! + \! \frac{9 }{4} z^{2}  \big{)}
\frac{m_{\pi}^{2}}{4 M^{2}} \! + \! \mathcal{O}(m_{\pi}^{4})  \Big{]} \, .  \;\;\;\;
\end{eqnarray}
where $ z = \frac{ \Lambda^{4} M^{2}}{\pi^{2} (\Lambda^{2} + M^{2} )^{2} f_{\pi}^{2}} $,
$M$ is the constituent quark mass, $m_{\pi}$ is the pion mass, $f_{\pi}$ is the pion decay constant,
and $\Lambda$ the cutoff in the NJL model.

\begin{figure}[htb]
\centering
\includegraphics[width=0.45\textwidth]{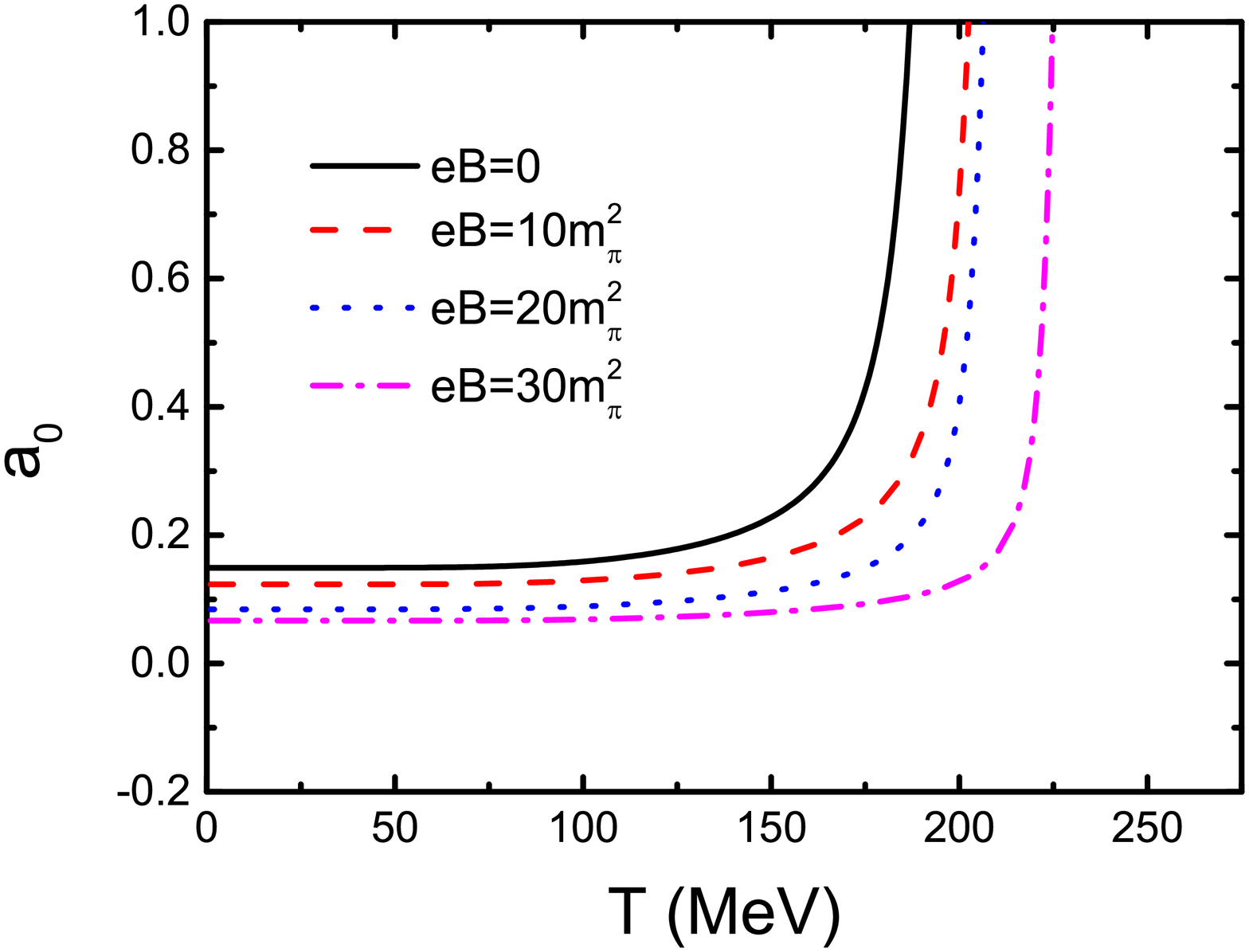}
\includegraphics[width=0.45\textwidth]{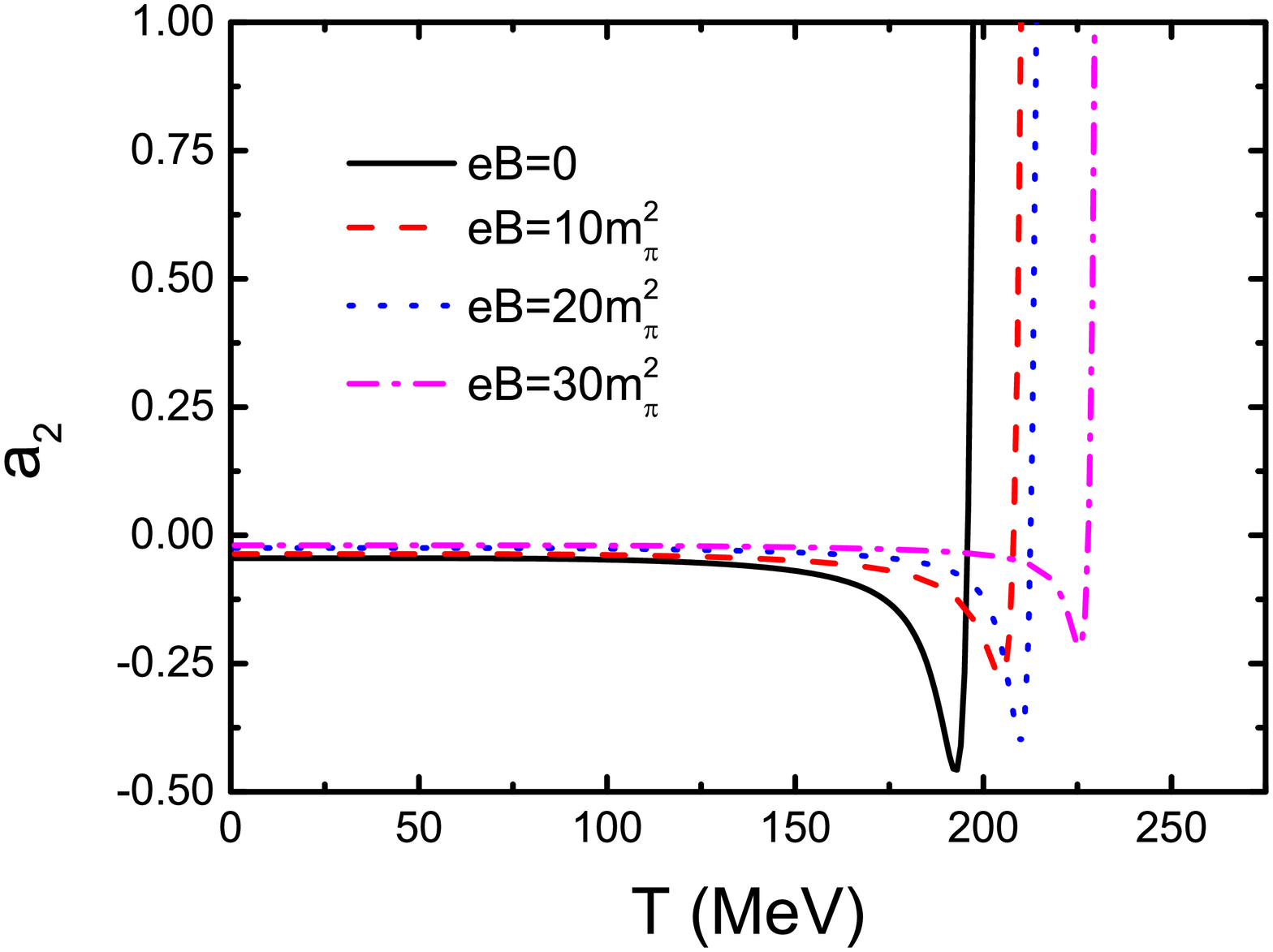}
\caption{(color online) Calculated scattering lengths $a_{0}^{}$ and $a_{2}^{}$ as functions of temperature in  cases of zero and several nonzero magnetic field strengths.}\label{fnew2}
\end{figure}

It is apparent that the above relation can be extended to the case at finite temperature and finite magnetic field with only taking the $M$, $m_{\pi}$, $f_{\pi}^{}$ and $\Lambda$ in case of finite temperature and finite magnetic field as the inputs.
With those obtained in last section as the inputs we get the scattering lengths
in case of vanishing and nonzero magnetic field strength.
The obtained results are shown in Fig.~\ref{fnew2}. The figure manifests evidently that, in both the cases of zero and nonzero magnetic field strength, the $a_{0}^{}$ and $a_{2}^{}$ all keep corresponding constant in low temperature region.
With increasing the magnetic field strength, the absolute value of the ``constant" gets smaller.  Whereas as the temperature increases to the pseudo-critical temperature marked as $T_{c}^{r}$ in last section, both the $a_{0}^{}$ and the $a_{2}^{}$ diverge to positive infinity rapidly. Extending the discussion in Ref.~\cite{Fu:2009PRD}, such divergences mean that the pion may get melted at the temperature. In addition, combining such a feature with the meaning of $T_{c}^{r}$, we can infer that the dynamical chiral symmetry restoration and the quark deconfinement coincide with each other~\cite{Xin:2014PRD}.

\begin{figure}[htb]
\centering
\includegraphics[width=0.45\textwidth]{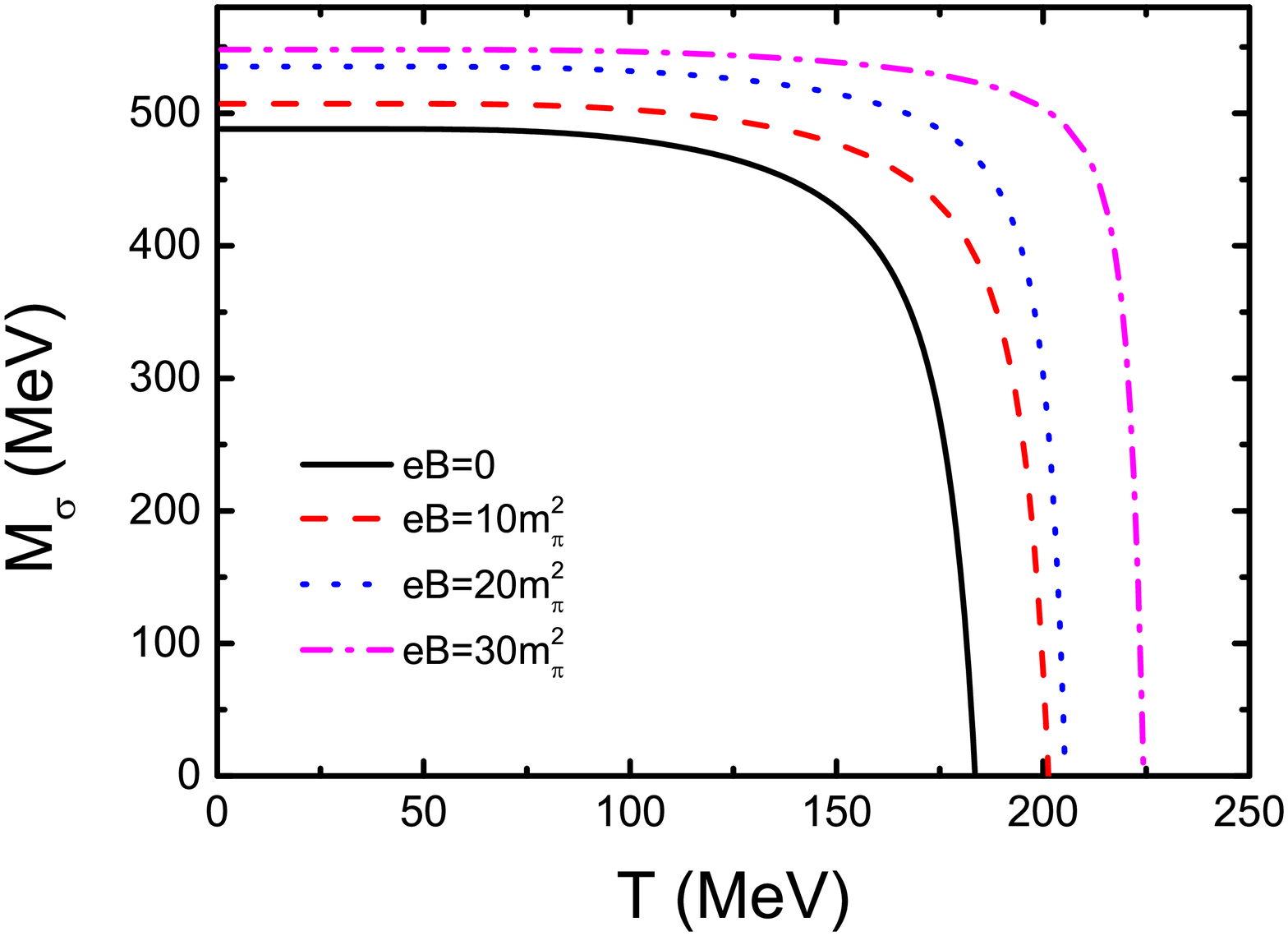}
\includegraphics[width=0.45\textwidth]{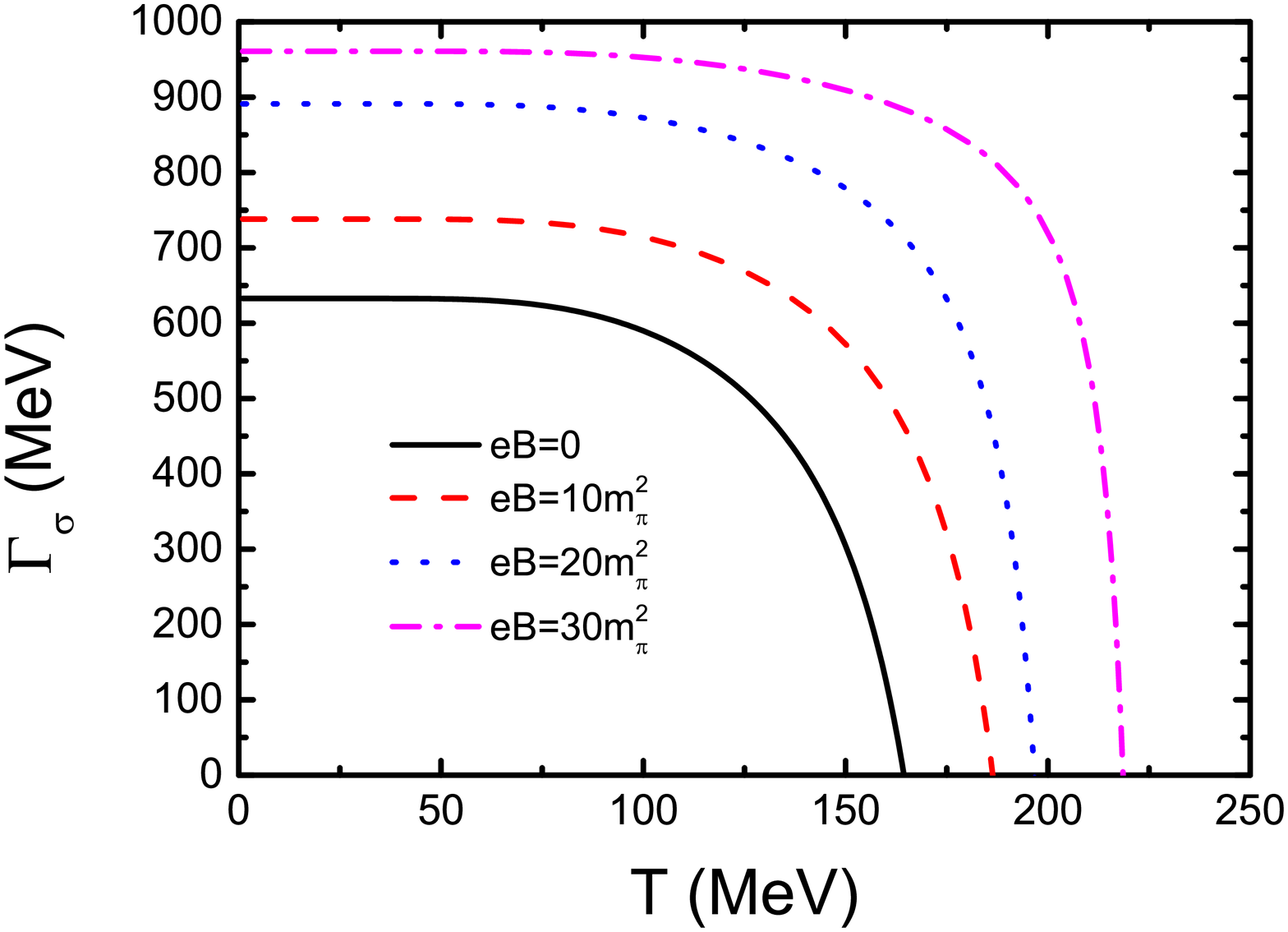}
\caption{(color online) Calculated $\sigma$ meson mass and its width as functions of temperature in  cases of zero and several nonzero magnetic field strengths.}\label{fnew4}
\end{figure}

Having had the $\pi$--$\pi$ scattering lengths at hand, one can obtain the masses and their widths of the $\sigma$ and $\rho$ mesons in the scheme of the Roy equation~\cite{Roy:1971PLB,He:1998NPA,Ananthanarayan:2001PR,Colangelo:2001NPB,Surovtsev:2008NPA,Caprini:2006PRL,Pelaez:2011PRL}. Following the method of Ref.~\cite{Caprini:2006PRL}, we can fix
the mass and width of $\sigma$ meson as
\begin{equation}
m_{\sigma}^{} = m_{0}^{} + m_{1}^{} \Delta a_{0}^{} + m_{2}^{} \Delta a_{2}^{} \, ,
\end{equation}
where
\begin{eqnarray}
\Delta a_{0}^{} & = & (a_{0}^{} - 0.22)/0.005\, ,  \label{eqn:Deltaa0}  \\[3mm]
\Delta a_{2}^{} & = & (a_{2}^{} +0.0444)/0.001 \, ,  \label{eqn:Deltaa2}
\end{eqnarray}
with $m_{0}^{} = 441 - i272$~MeV, $m_{1}^{} = -2.4 + i3.8$~MeV and $m_{2}^{} = 0.8 - i4.0$~MeV.
We can then have the temperature and magnetic field strength dependence of the $\sigma$ meson mass and width when the results illustrated in Fig.~\ref{fnew2} are taken as the inputs. The obtained results of the temperature dependence of the mass and the width at zero and several nonzero magnetic field strengths are shown in Fig.~\ref{fnew4}.
It shows obviously that, at zero temperature and zero magnetic field strength, the $\sigma$ meson mass $m_{\sigma}^{} = 488\;$MeV and its width $\Gamma_{\sigma}^{} = 633\;$MeV. They all agree with the data given in PDG~\cite{Nakamura:2010Review} and Refs.~\cite{Oller:2012PRD,Caprini:2006PRL,Pelaez:2011PRL} very well.
We can also see from the figure that the variation behavior of the mass in low temperature region is qualitatively consistent with the result we obtained by analyzing the internal quark--antiquark structure  described in last section except for that there exists a roughly factor 2 difference between the $m_{\sigma}^{}(T\!=\!0, eB\!=\!0)$.
The feature for the mass to decrease to $0$ but not increase at high temperature is due to the divergence of the scattering length.
Moreover, the decreasing characteristic of the width with respect to the temperature indicates
that such a scalar meson may not melt at high temperature, which is consistent with the lattice QCD result for heavy scalar mesons~\cite{Aarts:20113}.
The Fig.~\ref{fnew4} also manifests distinctly that the $\sigma$ meson mass increases with increasing  the magnetic field strength in low temperature region,
which is consistent with the result we obtained in last section.
Meanwhile the width of the $\sigma$ meson mass pole increases with the magnetic field strengths.

\begin{figure}[htb]
\centering
\includegraphics[width=0.45\textwidth]{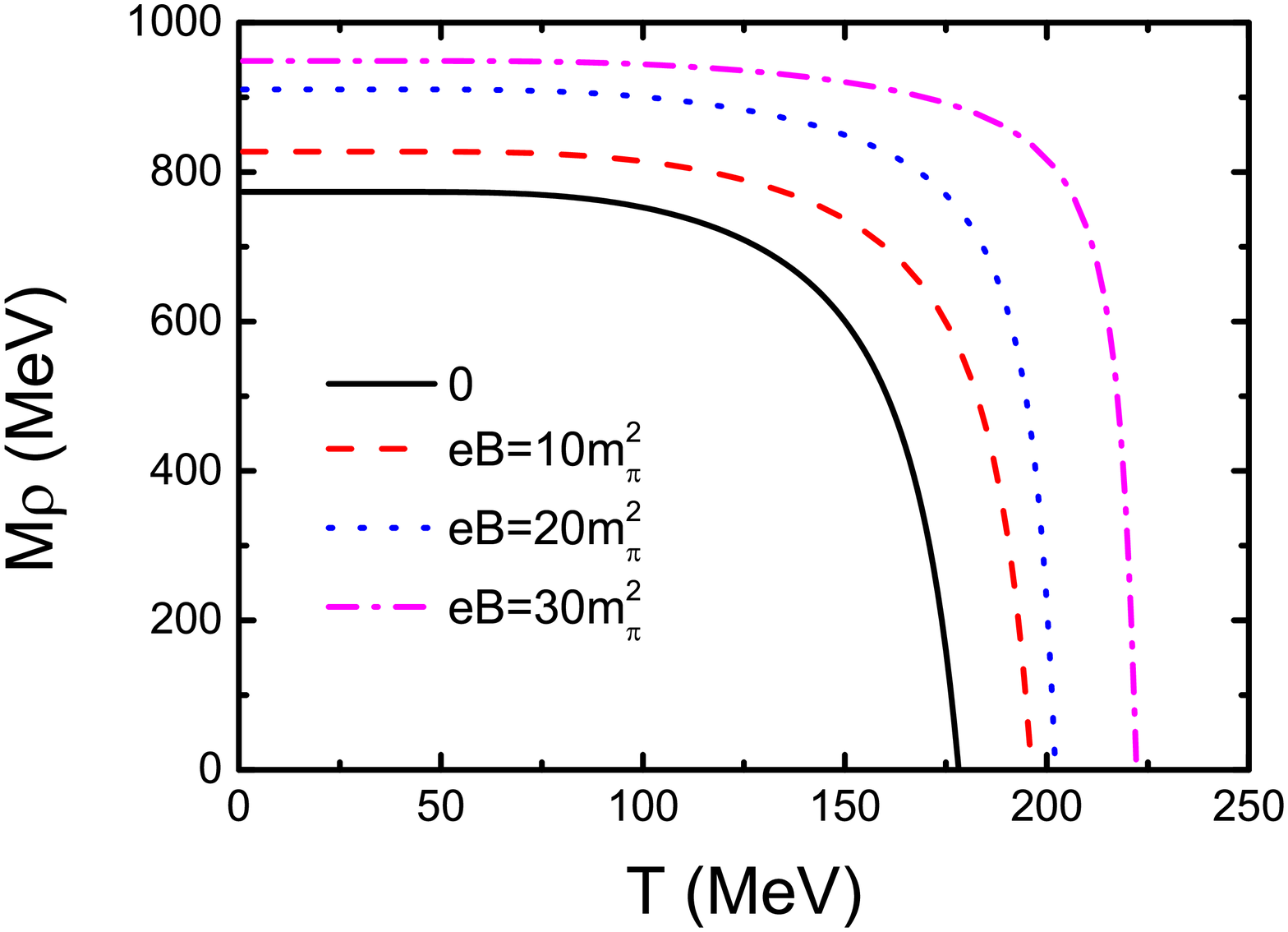}
\includegraphics[width=0.45\textwidth]{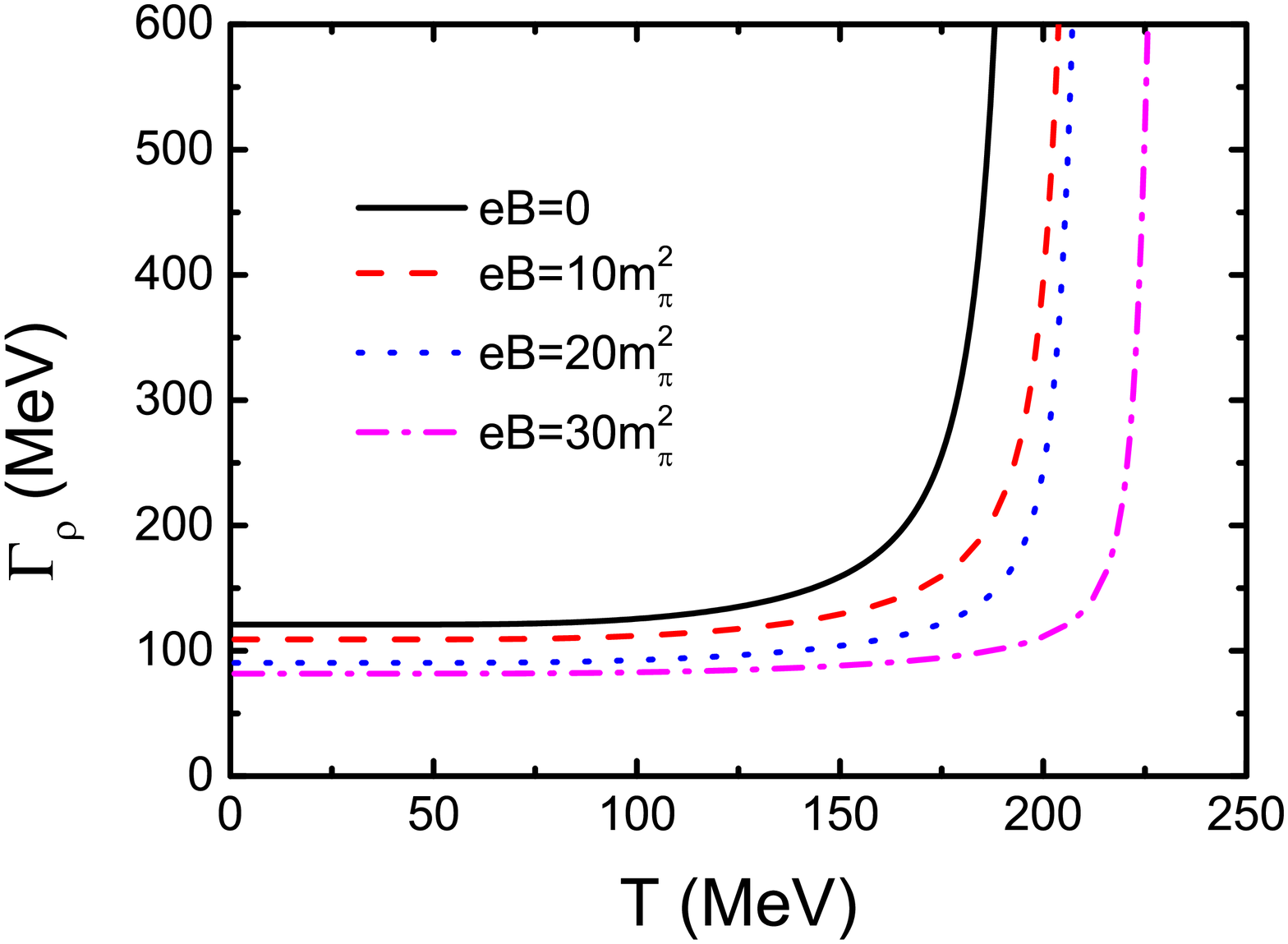}
\caption{(color online) Calculated $\rho$ meson mass and its width as functions of temperature in cases of zero and several nonzero magnetic field strengths.}\label{fnew6}
\end{figure}

In the similar way, we can also fit the $\rho$ meson mass and width in view of $\pi$--$\pi$ scattering
in the form
\begin{equation}
m_{\rho}^{} = m_{0,\rho}^{} + m_{1,\rho}^{} \Delta a_{0}^{} + m_{2,\rho}^{} \Delta a_{2}^{} \, ,
\end{equation}
with the $\Delta a_{0}^{}$ and $\Delta a_{2}^{}$ being the same as Eqs.~(\ref{eqn:Deltaa0}),  (\ref{eqn:Deltaa2}), respectively,
and parameters $m_{0,\rho}^{}=715.5-i73.5\,$MeV, $m_{1,\rho}^{} = -3.9 - i0.9\,$MeV and  $m_{2,\rho}^{} = 4.5 + i0.2\,$MeV.

\begin{table}[htb]
\caption{Calculated critical temperature for $\rho$ meson mass to be zero ($T_{c}^{\rho}$), the $\rho$ meson disassociation temperature ($T_{da}^{\rho}$) and comparison with the melting temperature ($T_{m}^{\rho}$) obtained in last section (All the temperatures are in unit MeV and the $eB$ in $m_{\pi}^{2}$ at zero temperature and zero magnetic filed). }
 \begin{tabular}
 {c|cccc} \hline \hline $eB$  & ~~$0 $~~ & ~~$10 $~~ & ~~$20 $~~ &   ~~$30 $~~ \\  \hline
$T_{c}^{\rho}$ &  $177$ &  $196$  &  $202$  &  $222$   \\
$T_{da}^{\rho}$ &  $187$ &  $204$  &  $207$  &  $226$   \\[1mm]
$T_{m}^{{\rho}^{0}}$   & $155$ &  $194$  &  $215$  &  $245$     \\
\hline \hline
\end{tabular}
\label{Critical-T-glance2}
\end{table}

 The obtained results of the temperature dependence of the $\rho$ meson mass and width at zero and several nonzero magnetic field strengths are shown in Fig.~\ref{fnew6}.
 It is apparent that the $\rho$ meson mass $774\,$MeV and width $121\,$MeV at zero temperature and zero magnetic field strength agree with the experimental data quoted in PDG excellently.
Meanwhile, in low temperature region, the $\rho$ meson mass almost maintains a constant,
and it decreases to zero quite rapidly as the temperature gets close to a critical value $T_{c}^{\rho}$.
With the increasing of the magnetic field strength, both the constant value and the critical temperature  ascend (some values of the $T_{c}^{\rho}$ are listed in Table~\ref{Critical-T-glance2}).
Such a feature is exactly the same that we obtained in last section.
Furthermore our calculations manifest that at any finite magnetic field strength, the width of the $\rho$ meson mass pole increases with the increasing of temperature; at a certain temperature, it decreases as the magnetic field gets stronger.
Especially the width diverges at certain critical temperature which increases as the magnetic field strength becomes larger. The divergence of the $\rho$ meson mass width means that
the life-time of the $\rho$ meson becomes zero, so that the $\rho$ meson will disassociate at the critical temperature. The obtained divergence temperature or the disassociation  temperature $T_{da}^{\rho}$ are listed in Table~\ref{Critical-T-glance2}. For comparison we re-quote the temperature for the $\rho_{0}^{}$ meson to melt, $T_{m}^{\rho_{0}^{}}$, in Table~\ref{Critical-T-glance2}.
The figure and the Table show obviously that the properties of the $\rho$ meson in view of the resonant state of pion-pion scattering are just the same as those obtained with analyzing the internal quark--antiquark structure of the mesons in last section.

\section{Summary and Remarks}

In this paper, we have calculated some properties of the scalar meson $\sigma$, pseudoscalar meson $\pi^{0,\pm}$ and vector meson $\rho^{0,\pm}$ at finite temperature and finite magnetic field in two distinct schemes in the NJL model.
One is the conventional that takes the mesons as quark and antiquark bound states.
Another is that regards the $\sigma$ and $\rho$ mesons as the pion--pion scattering resonant states.

%
%

To calculate the masses of the mesons sophisticatedly in the NJL model, we extend the $\Phi$--derivable method to finite magnetic field at first. Our calculation results manifest that the mass of the $\sigma$ meson in magnetic field keeps the same behavior as that in case of zero magnetic field, but with increasing magnetic field, the temperature dependence of the $\sigma$ meson mass becomes weaker. For the pseudoscalar meson $\pi$, the behavior becomes a little complicated. In a finite magnetic field, the neutral $\pi^{0}$ and the charged $\pi^{\pm}$ separate from each other, but they have similar dependence behaviors on the temperature, except for a slight quantitative difference. When temperature is lower than the critical value for nonzero magnetic field, the $\pi^{0}$ mass keeps almost a constant. Once the temperature reaches the critical value, the $\pi^{0}$ mass increases abruptly with the increase of the temperature, and becomes degenerate with the $\sigma$ meson mass. However the degeneracy is not precise because of the magnetic catalysis and the finite current quark mass effect. The charged $\pi^{\pm}$ mass increases with the magnetic field, no matter the temperature is lower or higher than the critical value. We also find that the critical temperature obtained from the $\pi$ mass is overall a little higher than that gained by analyzing the properties of the quark. Such a feature that different criteria give distinct critical temperatures implies that the chiral phase transition at finite temperature and finite magnetic field is a crossover.
For the vector meson, we also distinguish the neutral $\rho^0$ from the charged $\rho^{\pm}$ in our calculation. The obtained results display that the masses of not only the neutral but also the charged increase with the strengthening of the magnetic field at low temperature.
Whereas, at a certain magnetic field, the masses decrease generally with the increasing of temperature. When the temperature increases to the critical value, both the $\rho^{0}$ and $\rho^{\pm}$ mass solutions disappear, which implies that the vector mesons will melt. The melting temperature increases with the ascension of the magnetic field, and the $\rho^{0}$ melting temperature is slightly higher than that for $\rho^{\pm}$. Because the $\rho$ meson will melt at high temperature, there may not exist $\rho$-meson condensate in the QCD vacuum.

We have also calculated the temperature and the magnetic field strength dependence of the neutral pion decay constant and checked the GOR relation in case of finite temperature and finite magnetic field. Our calculation result of the decay constant agree very well with the previous one.
Meanwhile, we find that temperature influences the GOR relation more greatly than the magnetic field and the fluctuation of the ratio $r=\frac{f_{\pi}^{2} m_{\pi}^{2}}{-m_{0}^{} \langle\overline{q}q\rangle}$ can be a signal for the chiral phase transition.
Such an aspect shows again that the magnetic field preserves the DCSB.

To calculate and analyze the properties of the $\sigma$ and $\rho$ mesons in case of vanishing and nonzero magnetic field strength in the pion--pion scattering scenario, we take the formalism of the Roy equation by extending, for the first time, the calculation of the $\pi$--$\pi$ scattering lengths to finite magnetic field.  The masses and their widths at zero temperature and zero magnetic field strength we obtained agree with experimental data excellently. Our calculated results of the temperature and magnetic field strength dependence of the scattering lengths ($a^{0}$ and $a^{2}$) and the mass widths indicate that the $\pi$ meson and the $\rho$ meson will get disassociated at high temperature and strong magnetic field, and the disassociation temperature of each kind of the mesons is almost the same as the corresponding melting temperature obtained by analyzing the internal quark structure. Meanwhile increasing the magnetic field strength retards the disassociation. These features confirm that there does not exist (charged) vector meson condensate in the QCD vacuum at finite magnetic field.
Whereas the scalar meson $\sigma$ will not disassociate, which agrees with what lattice QCD calculation on heavy flavor mesons manifests.

All the obtained variation behaviors of the mesons' properties with respect to the temperature and the magnetic field strength provide further evidence for that the external magnetic field enlarges the dynamical chiral symmetry breaking area, i.e., there exists a magnetic catalysis. However it does not mean that we have reached an end, since we have not taken into account explicitly the magnetic inhibition~\cite{Fukushima:2013PRL}, the sphaleron~\cite{Chao:2013PRD} and other effects. Furthermore the NJL model is only a contact interaction approximation of strong interaction, which neglects the contributions of the complicated quark-gluon interaction vertex and the dressed gluon propagator.
Extending the result obtained in linear sigma model~\cite{Ayala:2015PRD}, one may infer that such a neglect should be the origin of the magnetic catalysis in the models.
In addition, we have not taken into account the temperature and magnetic field strength dependence of the cutoff in the calculations, either.
Then investigations on the meson properties with more sophisticated approaches ({\it e.g.}, the Dyson-Schwinger equation approach, incorporating  explicitly the magnetic field dependence of the quark-gluon interaction vertex and the gluon propagator, and so on) are necessary. On the other hand, the practical situation may, in fact, be more complicated, for instance, the effect of the magnetic field on the phase transition may depend on the field strength non-monotonically.

\bigskip

The work was supported by the National Natural Science Foundation of China under Contract Nos.\ 10935001, 11175004 and 11435001, and the National Key Basic Research Program of China under Contract Nos.\ G2013CB834400 and 2015CB856900.
W. J. Fu thanks also the support of Alexander von Humboldt Foundation
via a Research Fellowship for Postdoctorial Researchers.


\end{document}